%% file: MS.tex
\documentclass[fleqn,usenatbib]{mnras}
\usepackage{newtxtext,newtxmath} 
\usepackage[T1]{fontenc} 
\DeclareRobustCommand{\VAN}[3]{#2}
\let\VANthebibliography\thebibliography
\def\thebibliography{\DeclareRobustCommand{\VAN}[3]{##3}\VANthebibliography}

\usepackage{lipsum} 
\usepackage{xcolor}\usepackage{textgreek}\usepackage[utf8]{inputenc}\usepackage[english]{babel}
\usepackage{hyperref}
\hypersetup{colorlinks=true, linkcolor=blue,   filecolor=blue,  urlcolor=red, citecolor=blue,}
\usepackage{color,colortbl}\usepackage{tensind}\tensordelimiter{?} 
\usepackage{verbatim}\usepackage[normalem]{ulem}\usepackage{orcidlink}\usepackage{soul}
\graphicspath{ {./figs/} }

\input{define}
\def\red{}
  
\title[Sgr-B molecular-cloud complex]{Bow-shock structure of Sgr-B molecular-cloud complex in the Galactic Centre inferred from 3D CO-line kinematics} 

\author[Y. Sofue]{Yoshiaki Sofue\orcidlink{0000-0002-4268-6499}
\\ 
Institute of Astronomy, The University of Tokyo, Mitaka, Tokyo 181-0015, Japan \\  
E-mail: sofue@ioa.s.u-tokyo.ac.jp
}
 
\date{Accepted XXX. Received YYY; in original form ZZZ}
 
\begin{document}
\label{firstpage}
\pagerange{\pageref{firstpage}--\pageref{lastpage}}
\maketitle

\begin{abstract}
Three-dimensional (3D) bubble structure of the Sgr-B molecular-cloud complex is derived by a kinematical analysis of CO-line archival cube data of the Galactic Centre (GC) observed with the Nobeyama 45-m telescope. 
The line-of-sight depth is estimated by applying the face-on transformation method of radial velocity to the projected distance on the Galactic plane considering the Galactic rotation of the central molecular zone (CMZ). 
The 3D complex exhibits a conical-horn structure with the Sgr-B2 cloud located in the farthest end on the line of sight at radial velocity $v_{\rm lsr} \sim 70$ km s$^{-1}$, and the entire complex composes a lopsided bubble opening toward the Sun at $v_{\rm lsr}\sim 50$ to 30 km s$^{-1}$. 
The line-of-sight extent of the complex is $\sim 100$ pc according to the large velocity extent for several tens of km s$^{-1}$ from Sgr B2 to the outskirts. 
The entire complex exhibits a flattened conical bubble with full sizes $\sim 40 \ {\rm pc} \times 20 \ {\rm pc} \times 100 \ {\rm pc}$ in the $l$, $b$ and line-of-sight directions, respectively. 
Based on the 3D analysis, we propose a formation scenario of the giant molecular bubble structure due to a galactic bow shock, and suggest that the star formation in Sgr B2 was enhanced by dual-side compression (DSC) of the B2 cloud by the Galactic shock wave from up-stream and expanding HII region from the down-stream side of the GC Arm I in Galactic rotation.  
\end{abstract}


\section{Introduction}
\label{intro}

The Sgr-B molecular complex is the largest gaseous structure in the central molecular zone (CMZ) with total molecular-gas mass of a few $10^6 \Msun$, where a remarkable "hole-and-ring" structure around the radio sources Sgr B1 and B2 has been found 
\citep{1987ApJS...65...13B,bal88}.
The hole/ring structure has been interpreted in three ways:

(i) Molecular-gas ring left after local perturbation by the large mass concentration of the Sgr-B2 cloud \citep{bal88}.


(ii) Expanding molecular-gas bubble produced by the feedback of star-formation in Sgr B1 and B2 \citep{sof90}.


(iii) Cavity left after collision of the Sgr-B2 molecular cloud \citep{has94,sat00,eno22}. 

The first model that (i) the hole was made by the gravitational perturbation in the complex has not been examined since the discovery.
The second model, (ii) expanding-bubble and sequential star formation, was proposed as an alternative to (i) in our early paper \citep{sof90}, where we performed a detailed comparison of the spatial and kinematical distributions of molecular gas in the CO line and HII regions in radio continuum and recombination lines.
We have shown that the molecular hole composes a shell structure at LSR (local standard of rest) velocity $\vlsr \sim 45$ \kms associated with Sgr B1.
We proposed a model that an expanding-bubble around Sgr B1 hit the Sgr-B2 cloud at $\vlsr \sim 60$--70 \kms and triggered the recent and more active star formation in B2.
The star-formation (SF) activity in Sgr B2 further produced stellar winds and expanding shock waves, which caused the larger-scale bubble in the extended molecular outskirts.
More recent high-resolution CO-line mappings have shown that the complex is composed of a number of shell-like structures and high-density core clouds with large velocity extent \citep{oka98,tok19}.
Expanding shells and bubbles are more often recognized in the GC \citep{tsu97,tsu09,tsu15}.

The third idea in (iii) the cloud-collision paradigm has been extensively studied using molecular-line observational data, since the core-and-ring structure in the Sgr-B complex led to an idea that it may be a remnant of cloud collision \citep{has94,sat00}.
\citet{eno22} recently performed multi-wavelengths study and examined the model based on their criterion for the head-on collision of two clouds \citep{fuk21}. 
However, since the orbital elements of the two clouds, the necessary-and-sufficient condition for a head-on collision, have not been determined in the current papers reporting the "evidences" for collisions, we may still have room to consider an alternative scenario.

In this paper we explore the expanding-bubble model (ii) \citep{sof90} in more detail, and show that the Sgr B molecular complex can be reasonably explained by a galactic shock wave associated with an open cylindrical cavity caved from inside by an expanding HII region as the feedback of the star forming activity in Sgr B1 and B2. 
New aspect in this work is that we take into account the Galactic rotation of the CMZ in the analysis of longitude-velocity diagrams (LVD), which has been ignored in the current studies. 
The spatial and velocity extents of the molecular complex, $\sim 50\epc \times \sim 60 \ekms$, cannot be ignored compared to the radius $\sim 100 \epc$ and rotation speed $\sim 150 \ekms$ of the CMZ \citep{sof13}, because the rapid rotation with large differential velocity significantly affects the kinematics and evolution of a cloud with such a size.
The spread longitude-velocity feature associated with Sgr B complex manifests the Galactic rotation, leading to a largely extended distribution of the gas in the line-of-sight direction.

As to the basic structure of the CMZ \citep{sof22-3d,2023ASPC..534...83H}, we consider it to be a gaseous disc in Galactic rotation at $\vrot \sim 150 \ekms$ in the deep gravitational well of the Galaxy \citep{sof13}.
We consider that the molecular disc (CMZ) is superposed by spiral density waves, where the galactic shock-wave theory \citep{fuj68,rob69,rob72} applies to form dense molecular arms, GC Arm I and II \citep{sof95} .
In the galactic shock waves, supersonic flow in galactic rotation encounters a stagnated gas in the arm rotating at the pattern speed, causing high density compression along the arm.
In this paper, we assume that the GC Arm I and II compose the major structure of the CMZ, on which the SF regions Sgr B1/B2 and Sgr C are located \citep{sof90,sof95,sof22-3d}.

Here, Sgr B1 and B2 are the thermal radio sources containing HII regions with their 110$\alpha$-recombination-line coordinates at
Sgr B1: $(l,b,\vlsr, \Delta v)\sim (0\deg.51,-0\deg.05, +45.0 \ekms, 28.0$), and B2: ($0\deg.670, -0\deg.036, +64.6 \ekms, 41.0 \ekms$), where $\Delta v$ is the line width of the recombination line \citep{dow80}. 
The corresponding CO-line clouds are located at B1: $(l,b,\vlsr)\sim (0\deg.52,-0\deg.05,30-40 \ekms$) and B2: ($0\deg.66, -0\deg.04,60-70 \ekms$) as read in the maps in the following sections.

We consider a molecular-bow shock (MBS) in the spiral arm of CMZ, which is indeed observed in the Milky Way in the tangential direction of the 4-kpc molecular ring as W43 at G30+00 and at toward the Norma arm
 as G22 \citep{2019PASJ...71S...1S,2019PASJ...71..104S}. 
MBS is ubiquitous not only in the Galactic disc, but also in spiral galaxies and their central few hundred parsecs.
In fact numerous MBS and giant cometary HII regions (GCH) are found along the spiral arms of the barred spiral galaxy M83 \citep{sof18} based on the optical image of M83 taken with the Hubble Space Telescope \citep{2017AJ....154...51M}.
Recent mid-infrared images of nearby spiral galaxies with the James Webb Space Telescope (JWST) \citep{2023ApJ...944L..17L,2023ApJ...944L..22B} have revealed numerous dust "voids" in the face-on spiral galaxy M74, even in its central region, which are considered to be infrared views of MBS. 
It is stressed that the MBS + GCH structure is ubiquitous all over the galactic disc including the central region, which may also be the case in our Galactic Centre.
In this paper, we focus on the Sgr B molecular complex in  the CMZ of our Galaxy in analogy to such MBS in the central molecular-gas discs of nearby galaxies. 

In section \ref{sec2} we analyze the molecular gas distribution and kinematics in the \coth-line data cube.
Section \ref{sec3} is devoted to construction of three-dimensional (3D) molecular maps of the Sgr B complex by applying the face-on transformation (FOT) method to the rotating Galactic Centre disc.
In section \ref{sec4} we propose three possible mechanisms to explain the observed bubble property of Sgr B2 complex.
In section \ref{sec5} we discuss the result in comparison with the current models, and summarize the paper in section \ref{sec6}.
\red{We assume the distance to the Galactic Centre to be $R_0=8.2$ kpc \citep{2019A&A...625L..10G}}

\section{Molecular-cloud property of Sgr B complex}
\label{sec-two}
\label{sec2}

\subsection{Data} 
\red{
We used the archival data cube of the $^{13}$CO $(J=1-0)$ line emission at 110.27 GHz taken from the CMZ survey of $3\deg\times 0\deg.7$ region using the Nobeyama 45-m telescope \citep{tok19}.
The telescope had a FWHM (full width of half maximum) angular resolution of $15''$, velocity resolution of $1 \ekms$, rms (root mean squared) noise temperature $1$ K.
The data are re-gridded to a cube with $(7''.5 \times 7''.5 \times 2 \ekms)$ sampling with an effective resolution of $16.7''$. 
In the analysis we use the brightness temperature, $\Tb$, which is assumed to be equal to the radiation temperature, $T_{\rm R}$, used in the archival data, because the gaseous structures discussed in this paper are extended and well resolved by the beam. 
We also used partially the archival cube data of the HCN ($J=4-3$) line emission at 354.5 GHz from the GC survey with ASTE (Atacama Sub-mm Telescope Experiments) 10-m telescope \citep{2018ApJS..236...40T}. 
The telescope had angular resolution of $22''$, and the data are resampled into ($8''.5 \times 8''.5\times 2$ \kms) grids with the effective spatial resolution of $24''$.   
}

\subsection{Integrated intensity (moment 0) map: the mass and conversion factor}

\begin{figure*}   
\begin{center}    
\includegraphics[width=7.5cm]{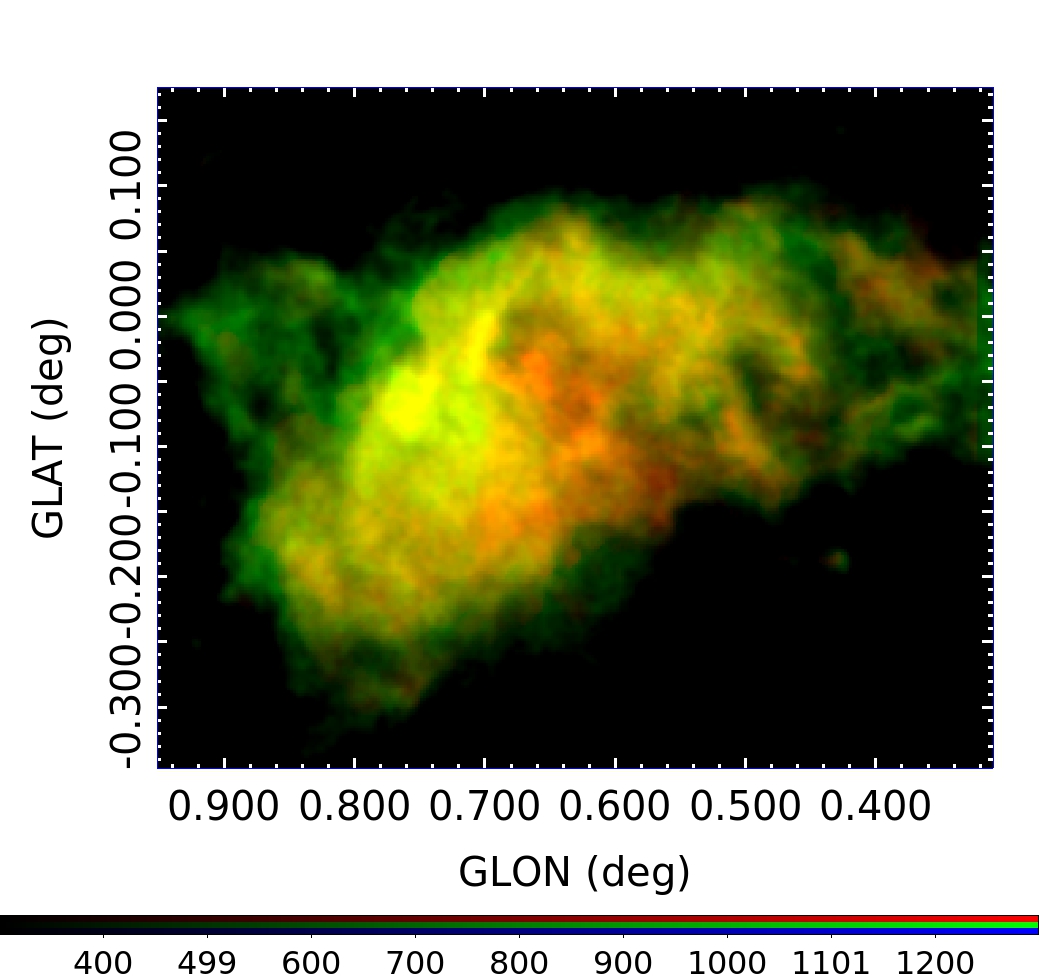} 
\includegraphics[width=7.5cm]{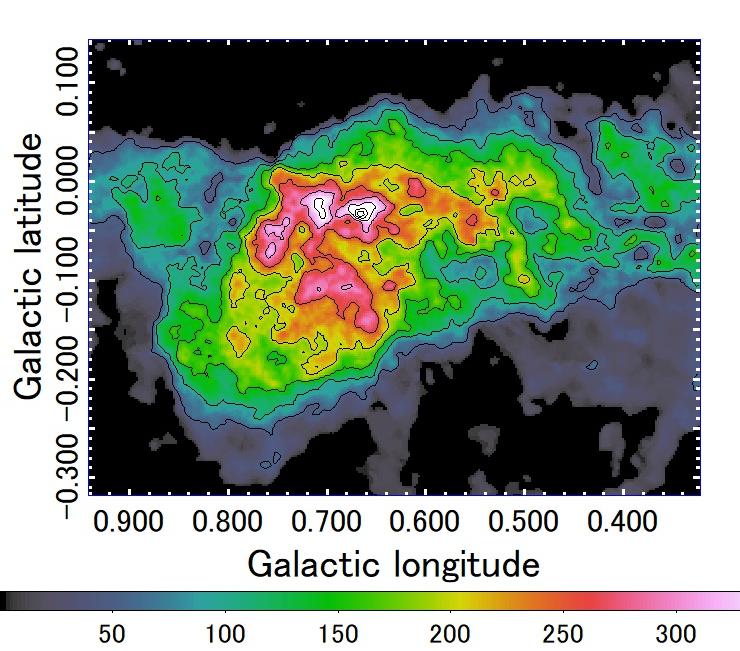}  
\includegraphics[width=7.5cm]{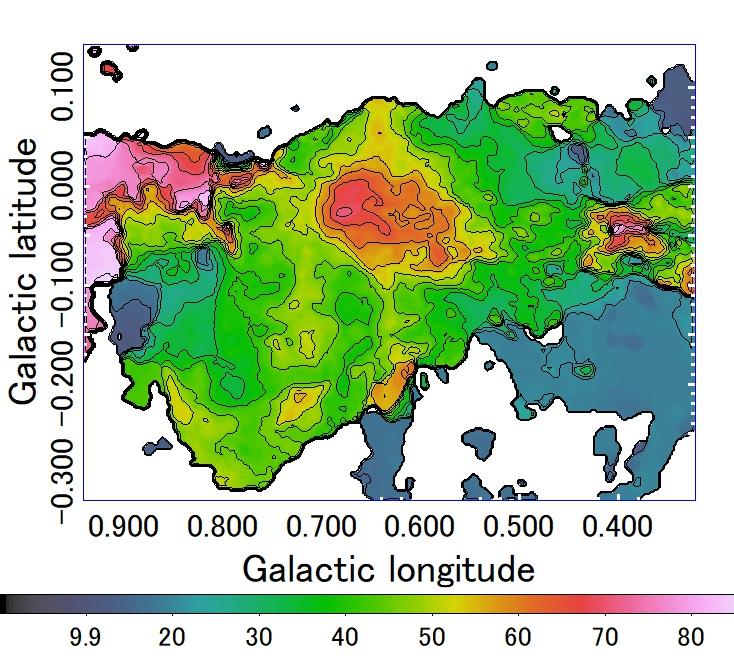}  
\includegraphics[width=7.5cm]{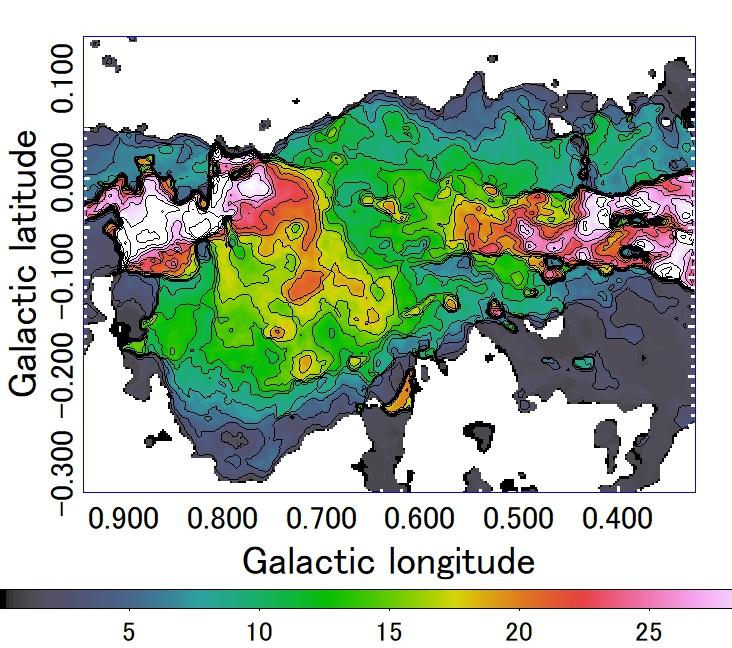}   
\end{center}
\caption{[Top left] Overlay of integrated intensity from $\vlsr=30$ to 70 \kms of the \co line emission in green and \coth in red of the Sgr B molecular cloud.  
[Top right] Moment 0, 
[bottom left] moment 1 and [bottom right] 
moment 2 in \coth line integrated from $\vlsr=10$ \kms to 90 \kms.
The horizontal broad feature is the contamination of Arm II at $\vlsr \sim 80$ \kms.
Note the difference from top-left panel, where the integration is from 0 to 70 \kms in order to avoid contamination of Arm II. 
}
\label{fig-moment}	 
\end{figure*}

Moment maps of the \co and \coth line intensities are shown in figure \ref{fig-moment}.
The major and minor axial diameters of the \co \red{moment 0} map are measured to be 
$D_x\sim 78.7$ pc and $D_y\sim 32.1$ pc, respectively, yielding the size radius of
$r=\sqrt{D_x D_y}/2\sim 25.1$ pc.
Adopting a CO-to-\Htwo conversion factor of $\Xco=0.5\times 10^{20}$ \xcounit and mean molecular weight $\mu=1.38$ \citep{ari96}, we have
$M_{\rm mol}=\pi r^2 \Ico \Xco (2 \mH \mu)\sim 1.7\times 10^6\Msun$.

The velocity width is measured to be $w_1\sim 12 \ekms$ on the moment 2 map, which is nearly uniform over the cloud.
The velocity gradient observed in the moment 1 map adds another velocity width due to internal motion by rotation, expansion and/or contraction of 
$w_2=\sqrt{(v_{\rm max}-v_{\rm min})^2}\sim 10$ \kms, where  
$v_{\rm max}\sim 52$ and $v_{\rm min}\sim 32$ \kms as read on the moment 1 map.
Thus we obtain an effective velocity dispersion of 
$v_\sigma=\sqrt{w_1^2+w_2^2}\sim 15.6$ \kms.
The dynamical (virial) mass is then estimated by
$M_{\rm vir}=r v_\sigma^2/G \sim 1.4\times 10^6 \Msun$.
The corresponding kinetic energy is 
$E_{\rm k}\sim 1/2 M_{\rm vir} v_\sigma ^2 \sim 3.4\times 10^{51}$ erg.

Therefore, the molecular mass is about 1.24 times the dynamical mass, and the cloud is bound by the gravity.
However, if we assume that the cloud is virialized, or the gravitational force is balanced with the dynamical pressure by the velocity dispersion, so that $M_{\rm vir}=M_{\rm mol}$, we can determined the conversion factor to be 
$\Xco\sim 0.4\times 10^{20}$ \xcounit.
This value is consistent with that predicted by the gradient of $\Xco$ with the galacto-centric distance as found for general values in the disc \citep{ari96}.

\begin{figure*}   
\begin{center}     
\hskip -2mm
\includegraphics[width=18cm]{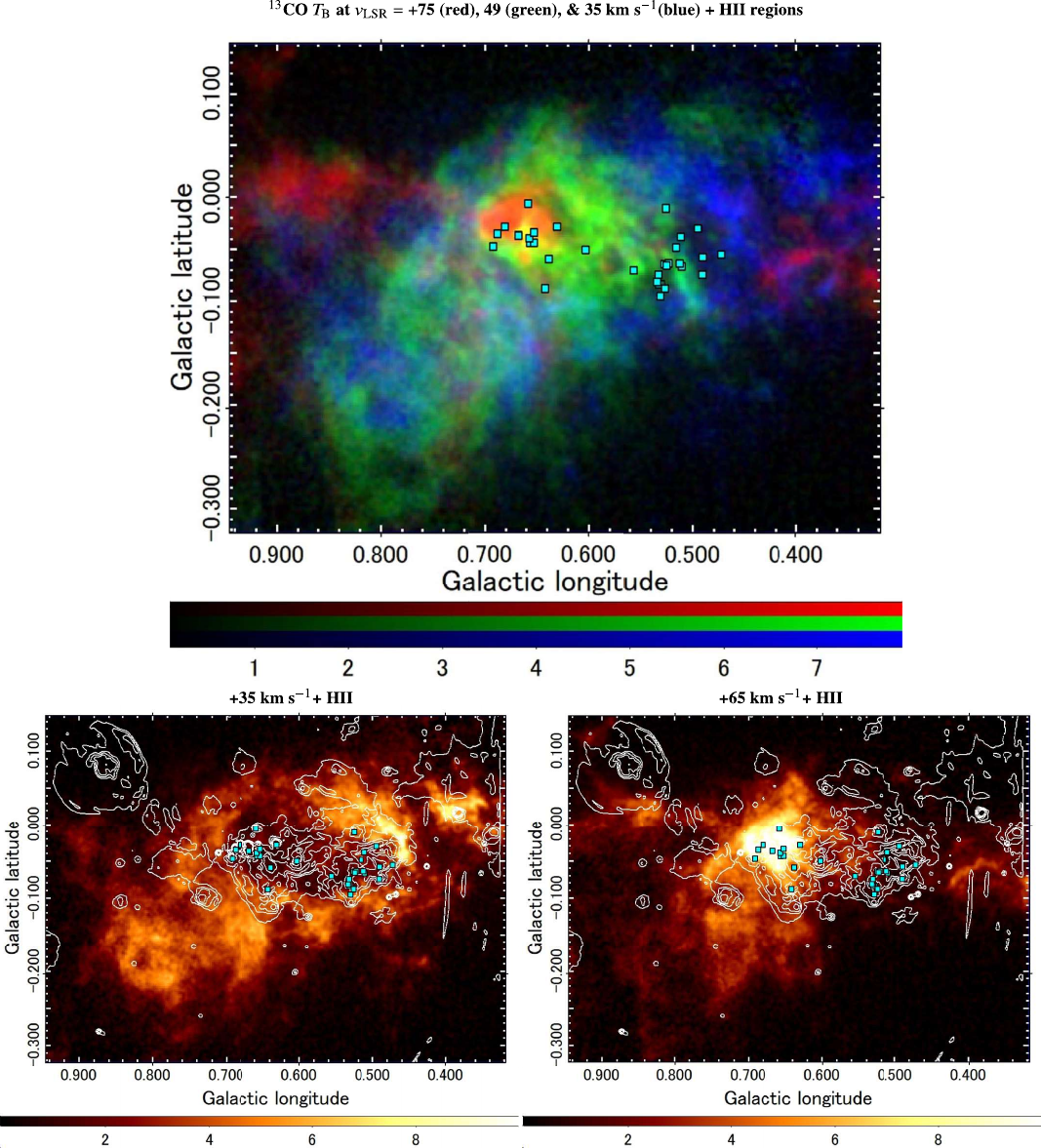}\\%
\end{center}
\caption{Overlay map of the brightness temperature $\Tb$ (scaled by the bars in K) at $\vlsr= 35$ (in blue), 49 (green), and 75 \kms (red).
Rectangles are positions of HII regions detected in the 110$\alpha$ recombination line \citep{meh92,meh93}.
The cluster at $l\sim 0\deg.67$ toward the red molecular clump is the Sgr B2 HII region, and that at $l\sim 0\deg.51$ is Sgr B1.
[Bottom] Same, but at $\vlsr=+35$ (left: as blue in the top panel) and 65  \kms (right: as red).
The contours are 1.3 GHz continuum surface brightness by MeerKAT \citep{hey22,yus22a}.
Rectangles are the same as the top panel.
}
\label{fig-rgb}	
\end{figure*}

\subsection{Velocity field and dispersion (moment 1 and 2) maps showing conical outflow toward the observer}

Moment 1 map, or the velocity field, is shown in figure \ref{fig-moment}, which reveals  negative velocity gradients from the center toward the edge of the cloud.
The radial velocity decreases from $\vlsr\sim 70 \ekms$ in the central red clump toward the eastern edge at $\sim 40$ \kms (green outskirt) at velocity gradient of $dv/dx\sim 0.4$ \kms pc$^{-1}$, and at $dv/dx\sim 0.8$ \kms pc$^{-1}$ toward the western edge at $30$ \kms (blue outskirt).

Moment 2 map shows that the dispersion velocity attains minimum of $\sigma_v\sim 8$ \kms in the central molecular clump that is surrounded by an extended cloud with larger velocity dispersion of $\sigma_v\sim 12$ \kms.
These velocity structures can be interpreted as due to an accelerating conical gas flow in the direction of the observer (Sun), blue-shifting from the central red clump toward the outer envelope.
This structure can be more clearly demonstrated by a composite color map shown in figure \ref{fig-rgb}, where the three different-velocity clumps are overlaid with $\vlsr=71$ \kms in red color, $51$ in green, and $31$ \kms in blue.

\subsection{Channel maps: cavity, shell, and radio continuum sources}

\begin{figure*}   
\begin{center}   
\includegraphics[width=18cm]{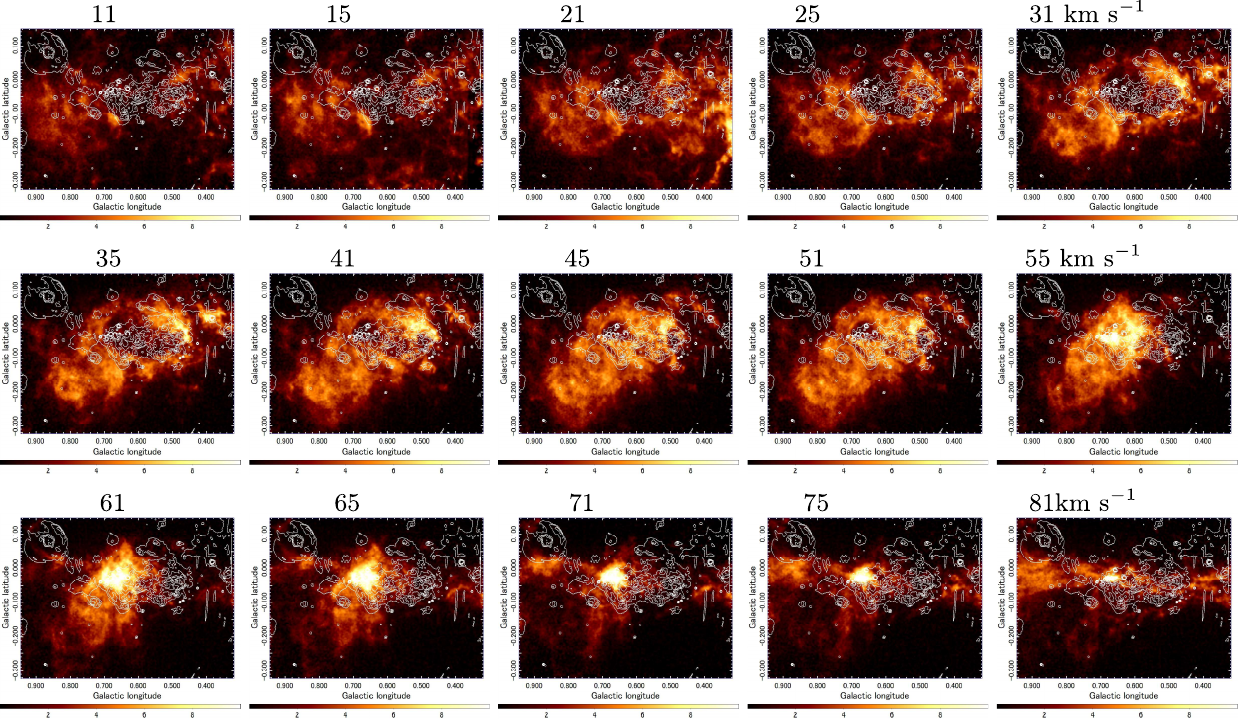}    
\end{center} 
\caption{Channel maps of \coth $\Tb$ in K at $\vlsr= 11$, ....., 81 \kms in order to demonstrate that the cavity surrounded by a shell (bubble) and the central Sgr B2 cloud are continuous structure in the velocity direction, showing no signature of distinct two clouds, as the cloud collision model postulates. } 
\label{fig-chan+mkat}	
\end{figure*}

Figure \ref{fig-chan+mkat} shows channel maps of Sgr B molecular components overlaid with contours of the radio continuum emission at 1.28 GHz from the MeerKAT observations \citep{hey22,yus22a}.
The figure demonstrates the spatial correlation of the radio sources Sgr B1 and B2 with the molecular cavity in the  extended components of Sgr B molecular cloud.

Figure \ref{fig-rgb} shows overlay of the brightness at 75 \kms in red, 40 \kms in green and 35 \kms in blue. 
The red clump near the cloud center composes a compact and dense central clump associated with Sgr B2, and is embedded in the green and blue components widely spread around the center clump.
As a whole, the Sgr B complex composes a conical horn structure opening from the central compact red clump in the GC-Arm I toward the green and blue-shifted outskirts. 

The radio continuum source Sgr B1 associated with a supernova remnant is located in touch with the western wall of the cone making the green-component of the extended cloud in figure \ref{fig-rgb}.
It is stressed that the western wall is observed as an arc-shaped high-density molecular arc showing up at $\sim 30$--50 \kms as shown in the upper panel of figure \ref{fig-chan+mkat}.

The Sgr B2 radio continuum source is full of point-like sources, mostly compact HII regions, surrounded by extended HII diffuse emission, and is located in the highest-density molecular clump recognized as the red component, as shown in  figure \ref{fig-chan+mkat}.
 
\begin{figure*}   
\begin{center}      
\includegraphics[width=8.6cm,height=10cm]{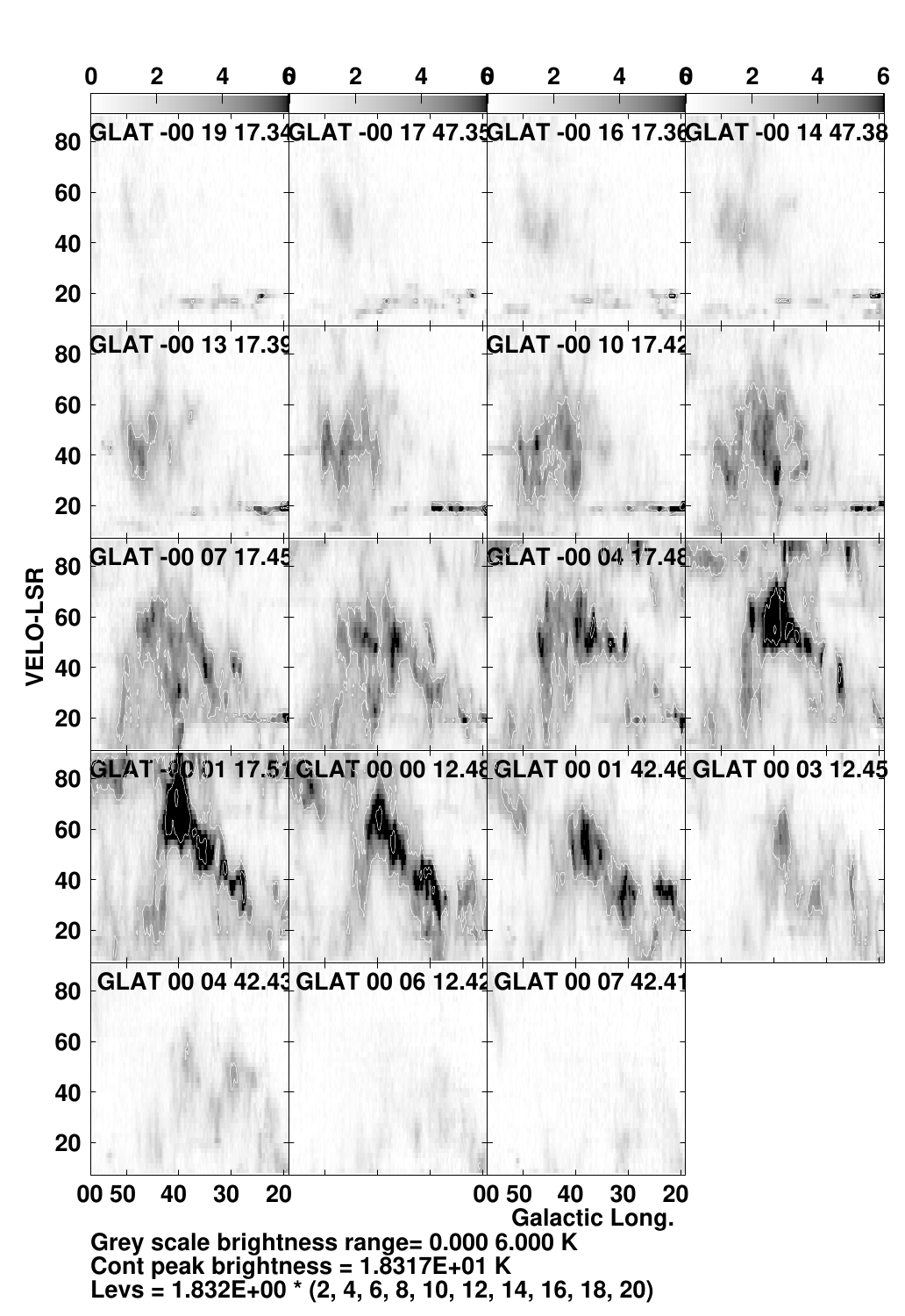} 
\includegraphics[width=8.6cm,height=10cm]{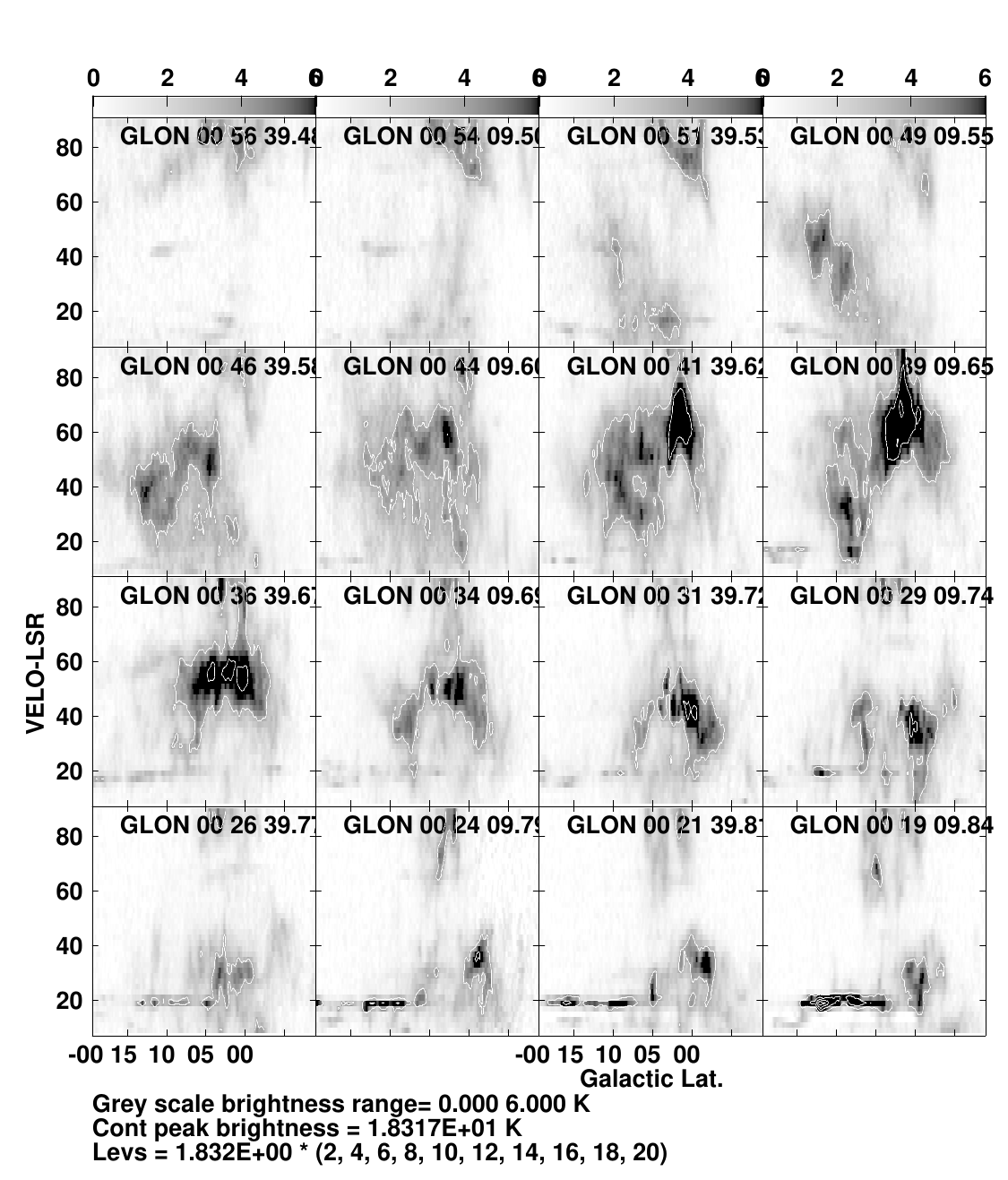}  
\end{center}
\caption{[left] LVD (Longitude-velocity diagrams) at various $b$.
[Right] BVD at various $l$. 
}
\label{fig-lvchan}	
\end{figure*} 

\subsection{\red{LVDs, BVDs, and kinematics}}

\begin{figure*}   
\begin{center}       
(A) \coth ($J=1-0$) LVD\\
\includegraphics[width=10cm]{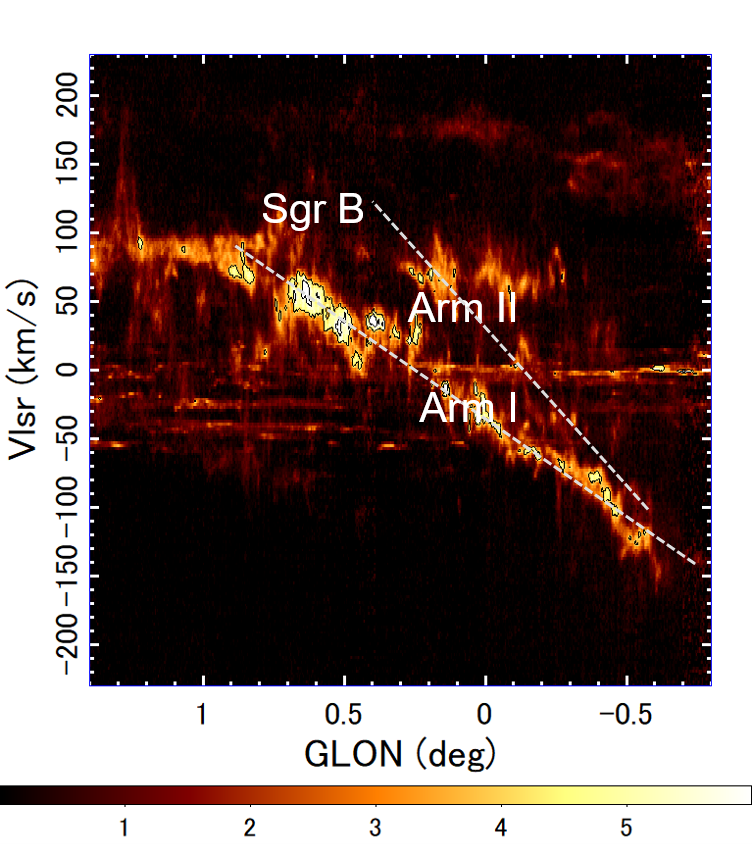}   \\
\red{\bf 
\hskip 1cm
(B) \coth LVD 
\hskip 3.5cm 
(C) \coth BVD
\hskip 3.5cm 
(D) HCN($J=4-3$) BVD} \\ 
\includegraphics[width=12.3cm,height=7.1cm]{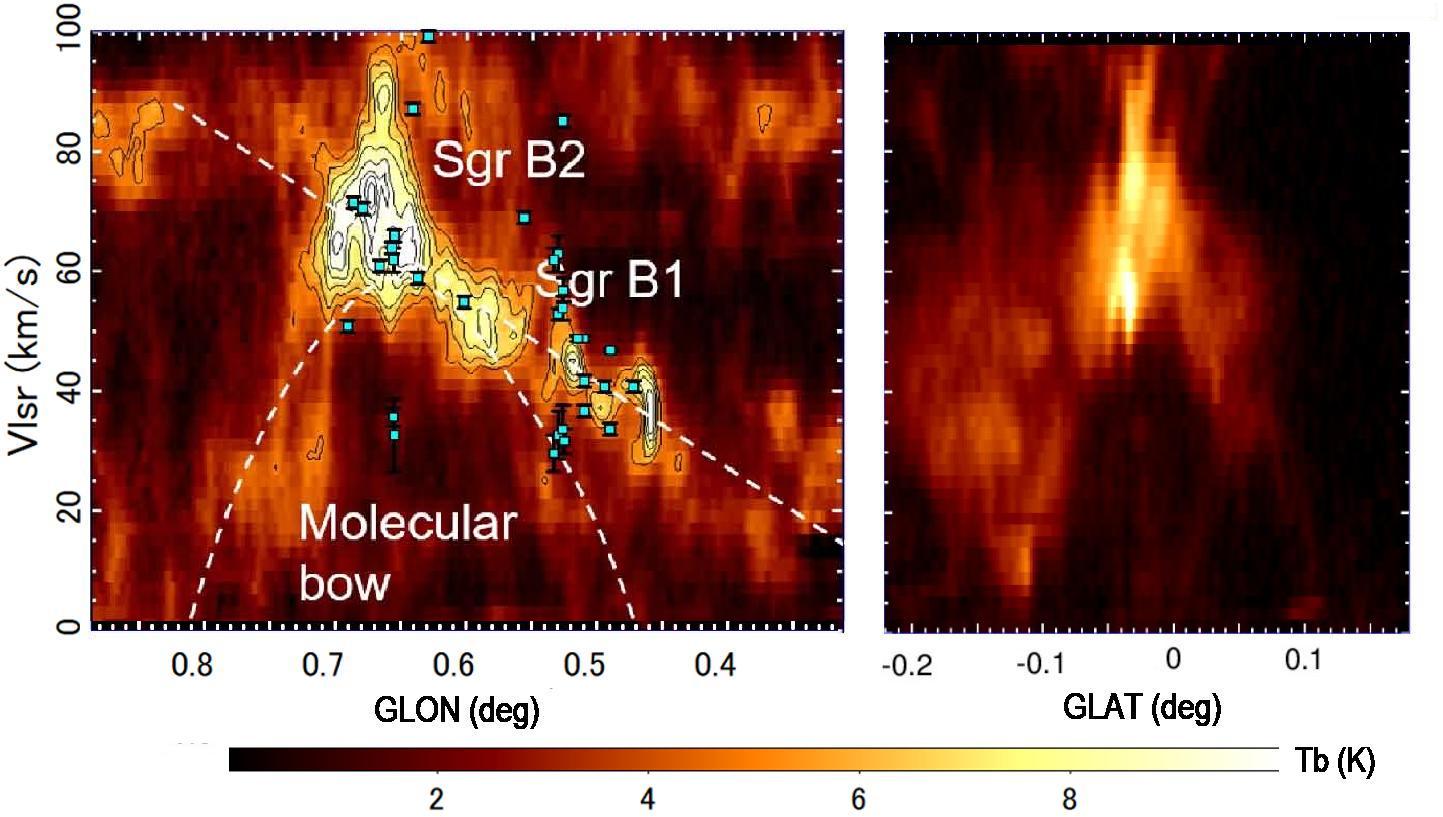}  
\hskip -5mm\includegraphics[width=5.2cm,height=6.9cm]{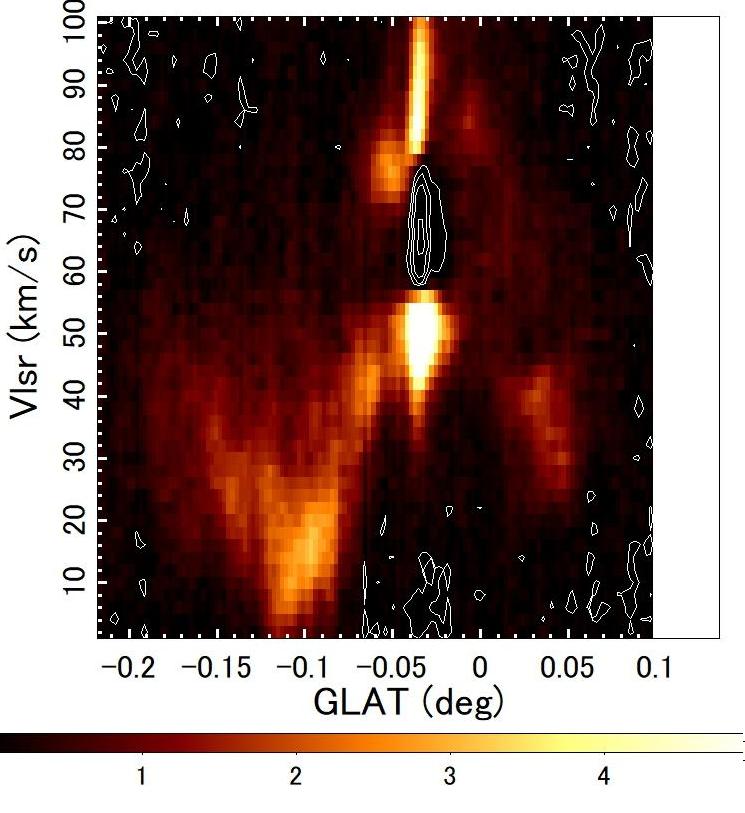} 

\end{center}
\caption{(A) \coth LVD along the entire ridge of GC Arm I at $b=+0\deg.02$.
Contours are every 1 K of $\Tb$ of the \coth line emission.
The Sgr-B cloud composes the major structure of the Arm, from which the 30--50 \kms cloud extends to compose a bow shape open toward decreasing velocity. 
(B) LVD across Sgr B2 at 
$(l,b,\vlsr)\sim (0\deg.66, -0\deg.04,65 \ekms)$ 
and Sgr B1 $(0\deg.52,-0\deg.05,35 \ekms$). 
Blue rectangles with error bars are H110$\alpha$ recombination-line sources of HII regions \citep{meh92,meh93}.
Sgr B1 and B2 and associated molecular clouds are located exactly on the LV ridge of the GC Arm I (straight dashed line), indicating that they are rotating with the Arm around the GC.
The curved dashed line indicates the extended outskirts at 20--50 \kms, composing a bow structure (conical cylinder) in the LVD.
\red{
(C) BVD perpendicular to the Galactic plane across Sgr B2, exhibiting a V ($\Lambda$) shape typical for a stagnated bow flow \citep{wil96} toward the down-stream (low velocity) side of the Galactic rotation.
Full width of the bow is $(\Delta X \times \Delta Z \times \Delta \vlsr) \sim (40 \epc \times 20 \epc \times 60 \ekms)$, \red{significantly narrower than the $l$-directional width in the left panel.
(D) BVD in the HCN ($J=4-3$) line, the high-density tracer of molecular density $\gtrsim 10^7$ H$_2$ cm$^{-3}$  \citep{2018ApJS..236...40T}. The gas is highly concentrated along the V-shaped bow shock with the peak at $\vlsr \sim 66$ \kms on the stagnation point. The line is saturated at Sgr B2 and exhibits absorption against the continuum of Sgr B2 as represented by white contours from $-0.1$ to $-2.1$ K every 0.5 K. Abscissa scales in the bottom three panels are approximately equal, showing that the vertical bow width is narrower than horizontal width.}}
}
\label{fig-lvd}	
\end{figure*}

The velocity structure is quantitatively represented by position-velocity diagrams (PVD).
Figure \ref{fig-lvchan} shows longitude-velocity diagrams (LVD) at various latitudes and latitude-velocity diagrams (BVD) at various longitudes of Sgr B complex.
Figure \ref{fig-lvd} shows LVD in wider longitudinal range, and enlargement near Sgr B complex.
The LVDs reveal that the main structure of GC Arm (GCA) I runs as a straight LV ridge running from $(l,v)\sim 0\deg.3,20\ \ekms)$ to $(0\deg.8,+100\ \ekms)$ as indicated by the dashed line in figure \ref{fig-lvd} (A).

The outskirts composed of the green (intermediate, $\sim 50$ \kms) and blue-shifted ($\sim 30$ \kms)  components of the complex are recognized as an open bow-shaped ridge in the LVD as traced by the dashed arc in panel (B) of figure \ref{fig-lvd}.
Thus, the molecular gas composing Sgr B1, B2 and the outskirts of the complex are smoothly connected on the LVD.
This fact indicates that they compose a continuous bow structure, or  there is no signature of discrete structures suggesting two different  clouds that are colliding at high speed.

Panel (C) shows the latitude-velocity diagram (BVD) across Sgr B2.
This BVD shows a clearer "V" shape, which is typical for a stagnated bow flow seen along the flow axis \citep{wil96}.
In the case of Sgr B2, the flow is toward the down-stream side of the Galactic shock wave at the GC Arm I.
\red{Panel (D) shows the same, but in the HCN $(J=4-3)$ line, as extracted from the survey of high-density tracers in the sub-mm wave lines in the GC with ASTE (Atacama Sub-mm Telescope Experiment) \citep{2018ApJS..236...40T}}.
The V-shaped bow features both in the LVD and BVD across Sgr B2 indicate a flattened corn structure in the $(l,b,\vlsr)$ space.
The full widths of the bow are measured to be $(\Delta X \times \Delta Z \times \Delta \vlsr) \sim (40 \epc \times 20 \epc \times 60 \ekms)$.
The V-shaped feature will be discussed later in more detail in a bow-shock model.

\subsection{Tight alignment of Sgr B1 and B2 clouds and HII regions along the GC Arm I}

HII regions detected in the 110$\alpha$ recombination-line are marked by blue rectangles on the $\Tb$ maps in figure \ref{fig-rgb} and on a LVD in figure \ref{fig-lvd} with error bars \citep{meh92,meh93}.

Recombination-line sources associated with Sgr B1 composes a cluster around $(l,b,\vlsr)\sim 0\deg.51,-0\deg.04,35\ekms)$ near the western edge of the molecular bubble at $\vlsr \sim 35\ekms$.
Those associated with Sgr B2 compose a cluster around $(0\deg.67,-0\deg.02,+65\ekms)$ tightly overlapped with the massive cloud (B2 cloud, or the red clump).

Sgr B1 and B2 molecular clouds and HII regions are  located exactly along the main LVD ridge of the GC Arm I. 
The exact alignment of the B1 and B2 clouds and HII regions in space and velocity with the Arm indicates that the molecular clouds nesting Sgr B1 and B2 are moving with the Arm without significant displacement from the Galactic rotation.

\section{Face-on transformation (FOT) and 3D structure}
\label{sec3}

\subsection{FOT}

In this section, we derive the three-dimensional (3D) distribution of molecular gas, assuming that the local velocities such as due to the expanding-bubble motion on the order $\sim 10 \ekms$ are sufficiently smaller than that of the galactic rotation on the order of $\sim 150 \ekms$. 
Because the rotation curve in the GC region at galacto-centric distances $R\sim 120 $ pc is nearly flat at $\vrot \sim 150$ \kms \citep{sof13}, we here assume that the gas in the CMZ is circularly rotating at a constant velocity of $150 \ \ekms$ 


The 3D position $(X,Y,Z)$ of a pixel with CO brightness temperature $T(v)$ at $(l,b)$ can be transformed from the observed position and velocity $(l,b,\vlsr)$ as
\be X=R_0 \ \sin \ l
\ee
\be Z=R_0 \ \tan  \ b, \ee
and  
\be
Y\simeq X \left(\frac{v}{\vrot}\right)^{-1} 
\sqrt{1-\left(\frac{v}{\vrot}\right)^2}.
\label{eqY}
\ee 

\subsection{3D structure}

\begin{figure*} 
\begin{center}  
\includegraphics[width=8cm]{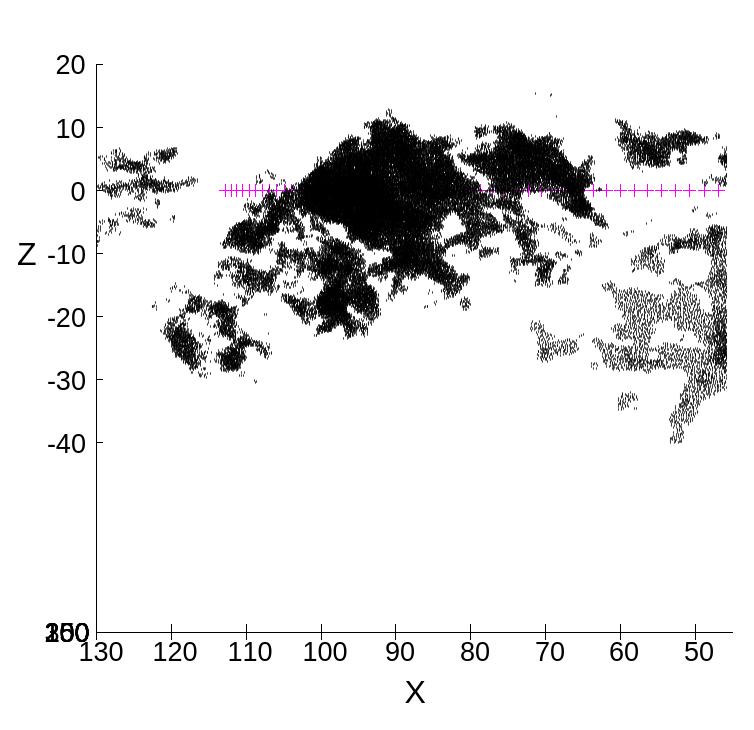}   
\includegraphics[width=8cm]{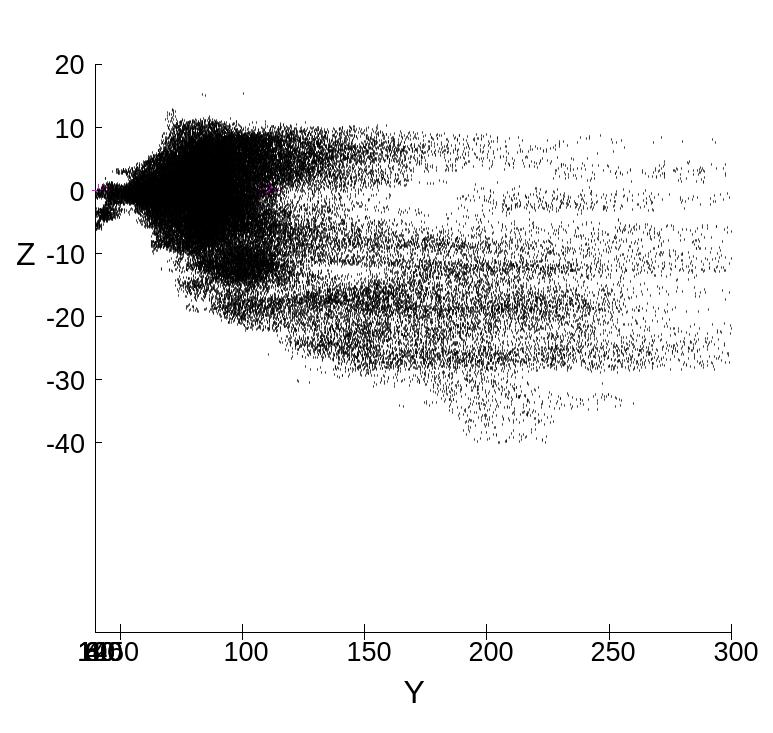}  \\
\includegraphics[width=8cm]{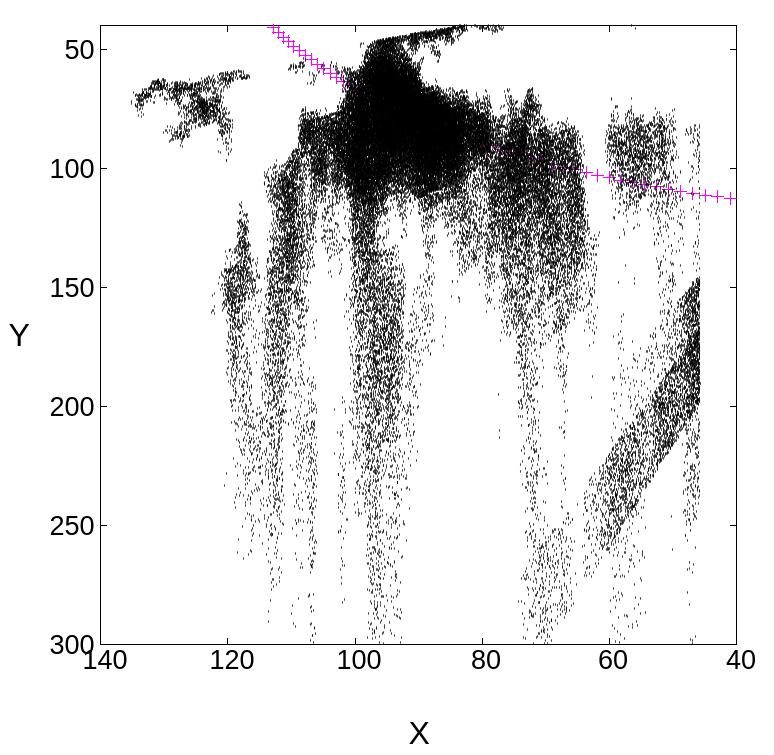} 
\includegraphics[width=8cm]{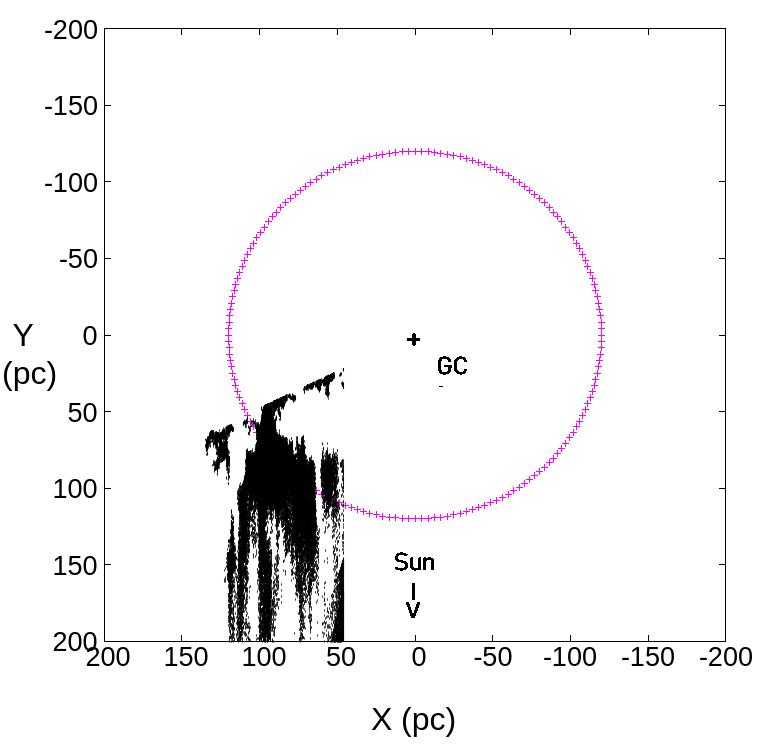} 
 \end{center}
\caption{[Top left] Density distribution of molecular gas in Sgr B molecular complex after applying the FOT projected on the sky ($(X,Z)$ plane) mimicking the moment 0 map. 
Scale is in pc.
[Top right] Same, but projected on the $(Y,Z)$ plane, showing a side view of the complex from the east, where $Y$ is increasing toward the Sun.
Note the bow shape about the Galactic plane. 
[Bottom left] Same, but projected on the Galactic plane ($(X,Y)$ plane). 
Magenta dots represent a ring in the galactic plane of radius 120 pc approximately representing the GC Arm I.
Thus obtained 3D gas distribution composes a molecular-bow structure sheathing the Sgr B2 SF region, making a giant cometary HII region (GCH) open toward the Sun.  
[Bottom right] Same, but on  wider $XY$ area including the CMZ. 
Sgr A$^*$ is at the center, $(X,Y)=(0,0)$, and the magenta circle is the 120 pc ring with GC Arm I and II \citep{sof95}.
The Sun is to positive $Y$ (downward)The outskirts of Sgr-B complex is extended for $\sim 100$ pc in the line-of-sight direction.}
\label{fig-3d}
\end{figure*}

\begin{figure*} 
\begin{center}    
\includegraphics[width=18cm]{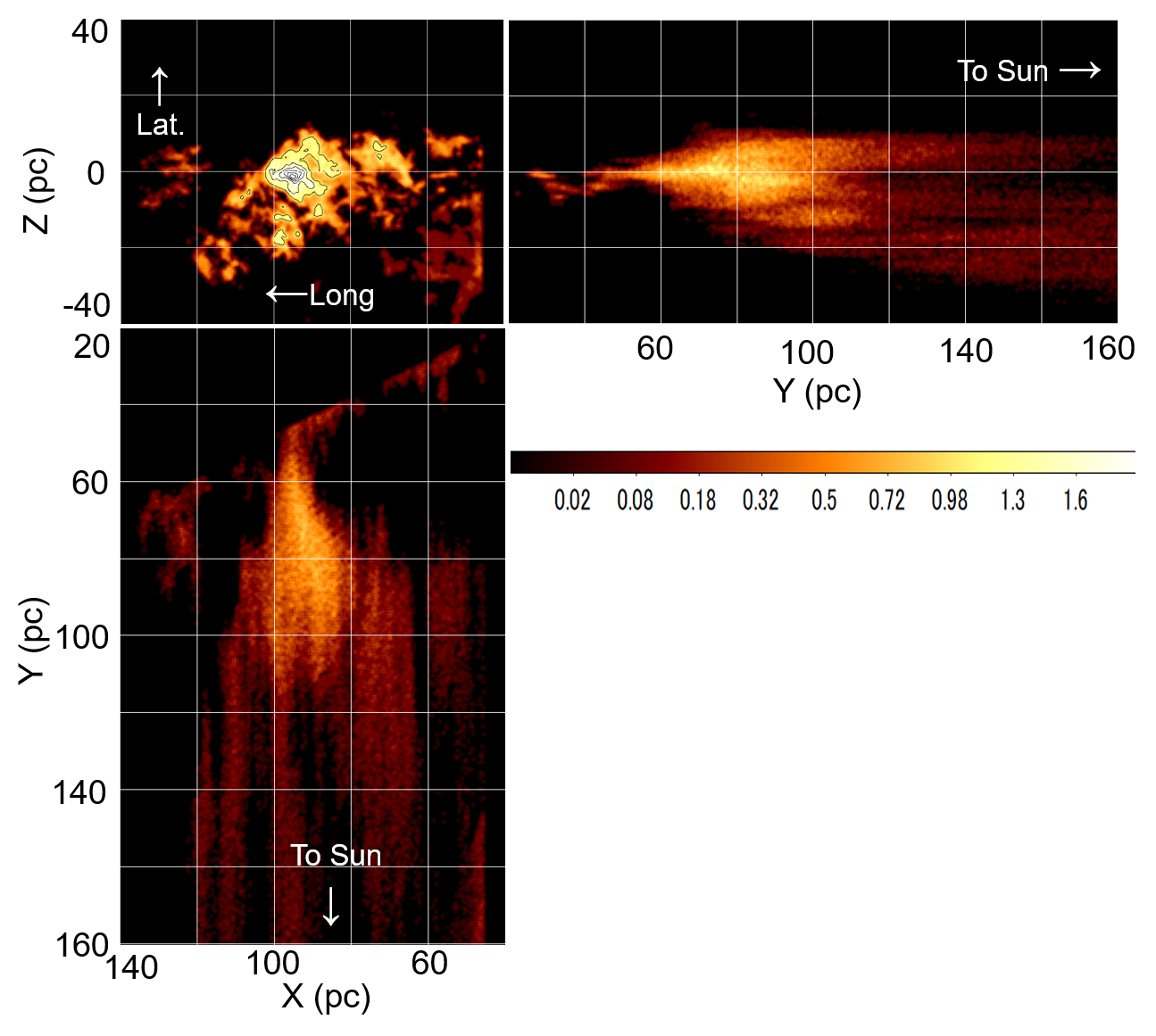}    
 \end{center}
\caption{Same as figure \ref{fig-3d}, but with the same linear scaling in order to compare the $X$, $Y$ and $Z$ extents in the same proportion. Intensity scaling is also common in all the panel in arbitrary unit. Contours in the top-left panel are drawn from 1 by interval 0.5. \red{Note that the $Y$- ($\vlsr$-) directional structure is continuous from the top to tail, indicating that the entire structure is a single object.}}
\label{fig-3d-xyz}
	\end{figure*}

Figures \ref{fig-3d} shows the result of FOT for the brightness clipping above $\Tb=2$ K.
In the figure the small dots are plotted so that the density of dots is proportional to the volume density of the molecular gas.
The top-left panel of figure \ref{fig-3d} is a projection of the gas-density distribution on the sky or on the $(X,Z)$ plane, representing the integrated column density, which is therefore identical to the moment 0 map of the CO line emission. 
The top-right panel is the projection on the $(Y,Z)$ plane, or a view from the right side (from eastern side). 
The bottom-left panel shows the projection on the $(X,Y)$ plane, or on the Galactic plane, so that it expresses the face-on view of the complex.
The magenta dots represents a ring of radius 120 pc in the galactic plane, approximately tracing the GC Arm I.
The bottom-right panel is the same, but projected on the $(X,Y)$ plane (galactic plane) in a wider area including the whole \red{giant molecular cloud (GMC)}.
Figure \ref{fig-3d-xyz} shows the same, but with the same linear scaling in the $X$, $Y$ and $Z$ directions with the intensity presented by grey scale in arbitrary unit as indicated by the bar. 

\begin{figure*} 
\begin{center}   
{\bf Stereogram}\\
{\bf $\odot$ Left eye \hskip 5.5cm $\odot$ Right eye}\\
\includegraphics[width=8cm]{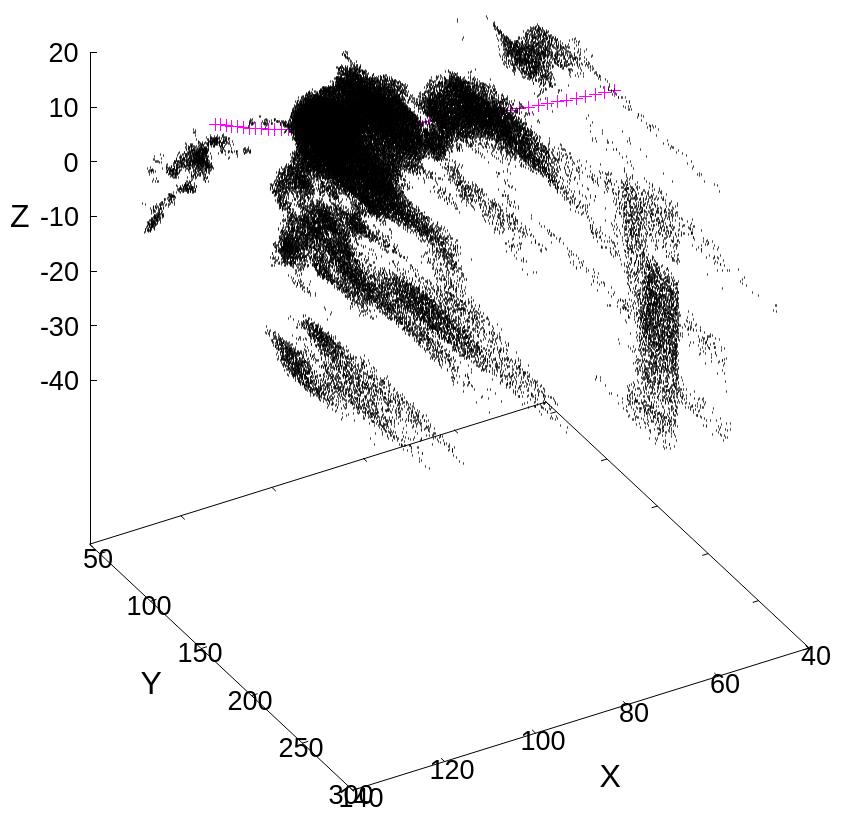}   
\includegraphics[width=8cm]{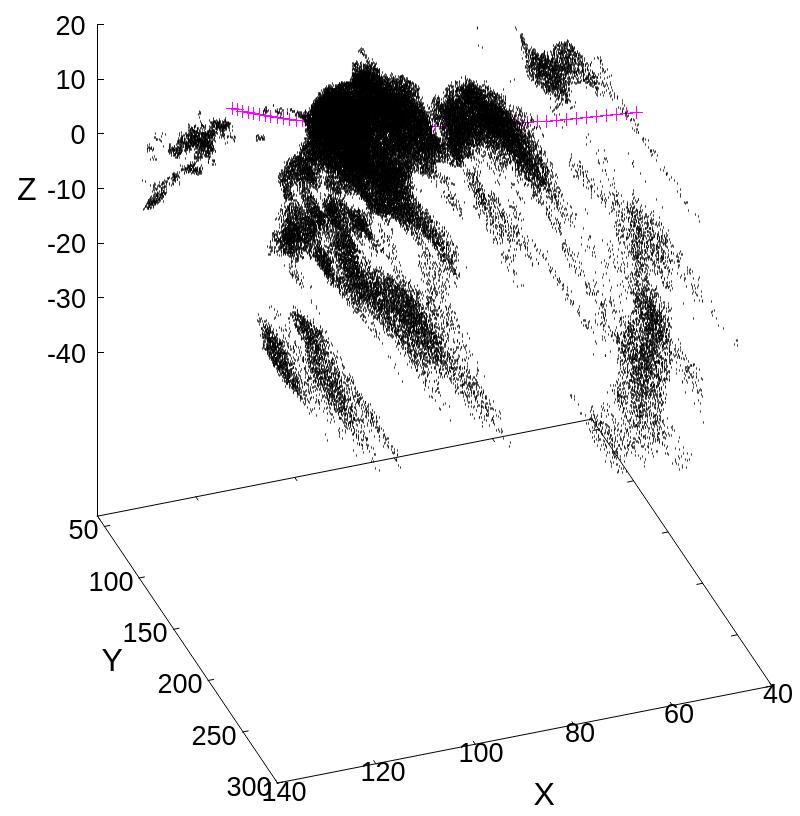}   
 \end{center}
\caption{Oblique projections of figure \ref{fig-3d} as seen from a standpoint in the north-eastern and nearer-side to the Sun.
Left and right panels show projections rotated by $30\deg$ around $X$ and $50\deg$ and $40\deg$ degrees around the $Z$ axis.
So, the two panels are seen from two directions by $10\deg$ rotated from each other, so that they can be used to obtain a stereo-graphic view of the gas distribution.  Magenta dots represent a ring in the galactic plane of radius 120 pc representing the GC Arm I.}
\label{fig-3d-stereo}
	\end{figure*}

Figure \ref{fig-3d-stereo} shows oblique projections of the plots viewed from a standpoint in the north-eastern and nearer side to the Sun.
The two (left and right) panels show the same, but projections from different standpoints rotated by 10 degrees from each other, so that the two panels can be used to obtain a stereo-graphic view of the gas distribution by viewing the panels by the two eyes.

\begin{figure*} 
\begin{center}     
\includegraphics[width=8.5cm]{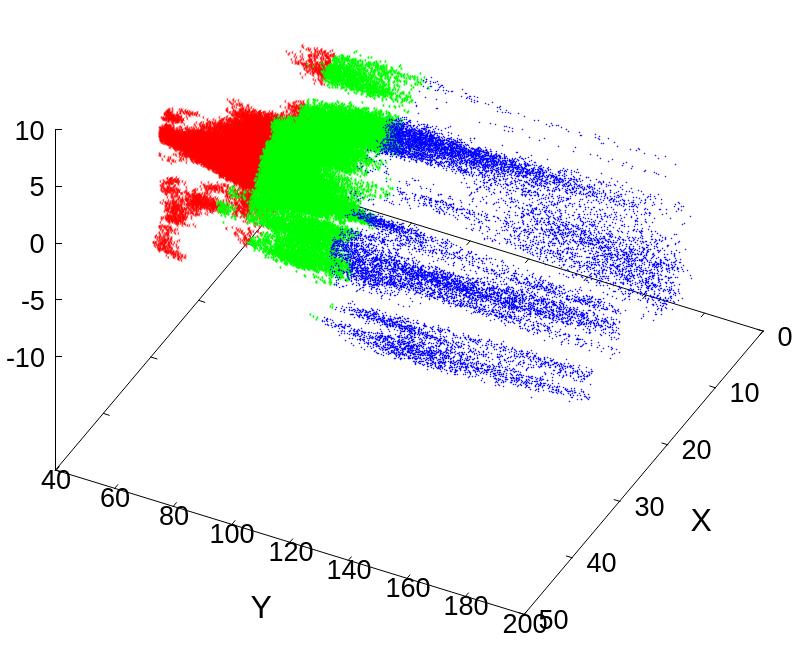}   
\includegraphics[width=8.5cm]{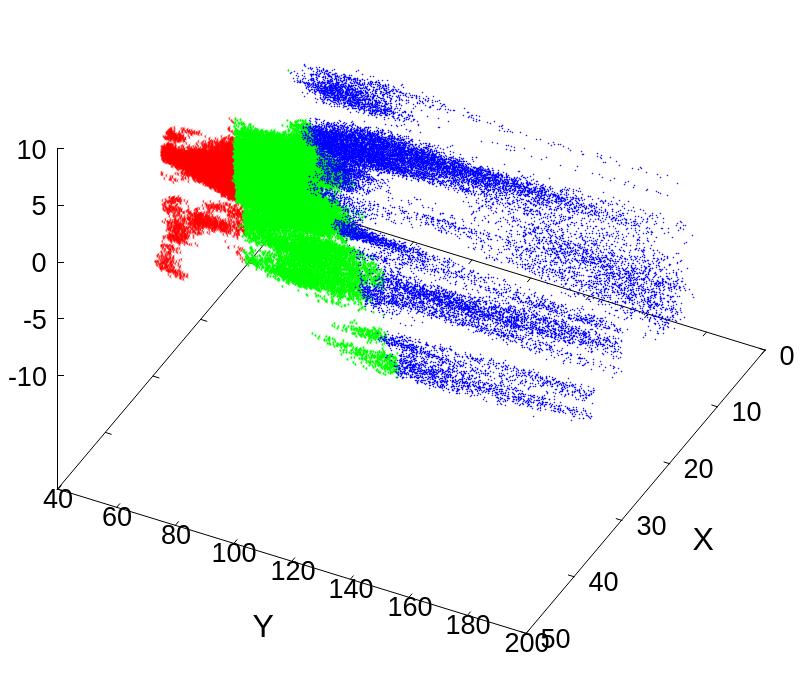}   
 \end{center}
\caption{[Left] 3D slices of figure \ref{fig-3d} projected on the sky ( $(X,Z)$ plane) at $Y<80$ pc by red, $80\le Y < 110$ in green, and $\ge 110$ in blue. 
[Right] Same, but sliced by velocity, showing large-, intermediate-, and low-velocity components in red, green and blue, respectively. }
\label{fig-slice}
	\end{figure*}

In order to see how the representative parts of the Sgr B complex according to the radial velocity corresponding to the resulting 3D distributions, we show in the left panel of figure \ref{fig-slice} the same plots, but giving three different colors red, green and blue (RGB) to the points having far, intermediate and near $Y$ values.
Namely, each colored regiong represents a slice perpendicular to the line of sight.
Similarly in the right panel of figure \ref{fig-slice} we show the same plots, but RGB colors represent points with high, intermediate, and low LSR velocities of the \coth-line emission. 
So, this diagram represents equal-velocity slices, each representing the 3D distribution of each of the RGB components in figure \ref{fig-rgb}.
 
Figures \ref{fig-3d} to \ref{fig-slice}, thus, show that the molecular gas is densest in the Sgr B2 clump at $\vlsr \sim 70 \ekms$, and the extended outskirts compose conical cylinder at velocities of $\sim 50$ to 30 \kms, which is open toward positive $Y$ direction (toward the Sun) and is blue shifted with respect to Sgr B2 densest clump.
It is stressed that the line-of-sight ($Y$) extent of the Sgr B complex, from Sgr B2 cloud to the blue-shifted outskirts, is as long as $\Delta Y \sim 100$ pc.
Namely, the complex composes a conical bubble of 3D ($X,Y,Z$) sizes of
\be
\Delta X \times \Delta Y \times \Delta Z \sim 40 \epc \times 100 \epc \times 20 \epc.
\ee
The line-of-sight size is longest, which is due to the large velocity width of $\Delta \vlsr \sim 60$ \kms.

\subsection{Uncertainty due to expanding motion and velocity offset}

\label{uncertainty}

In the present FOT we have ignored the expanding motion of the molecular gas.
If we correct for such local motion, the $Y$ directional extent of the 3D distribution is reduced, so that the bubble shape becomes more round and compact.
However, a possible velocity correction of $\sim 10 \ekms$ by the expansion of HII region is much smaller than the observed velocity extent of the entire Sgr B complex, $\sim 60 \ekms$, due to the Galactic rotation.
So, the correction by HII region expansion can reduce only $\sim 20$ percent at most in the $Y$ direction.
This means that the large velocity difference from $\vlsr \sim 20$ \kms to $\sim 70$ \kms in the Sgr B complex is due to the Galactic rotation, as simulated by a spiral structure formed in a disc in differential rotation \citep{kru19}. 
We comment that such stationary bubble/conical-horn structure may be similar to the dust voids observed in the spiral galaxies by JWST  \citep{2023ApJ...944L..17L,2023ApJ...944L..22B}. 

Another aspect to be mentioned is the uncertainty due to $\vlsr$ displacement from the rotation center on the LVD.
As equation \ref{eqY} indicates, the $Y$ directional distance depends on the radial velocity $v=\vlsr$ in the sense that the larger is $v$, the smaller becomes $Y$, and vise versa.
As shown in the LVD in figure \ref{fig-lvd}, the LV ridge (dashed line) representing Arm I intersects the rotation axis at $(l,\vlsr)\sim (0\deg, -30\ \ekms)$.
This implies that, if we correct for the displacement by $\delta v\sim -30 \ekms$, or replace $v$ by $v=\vlsr-\delta v$, the $Y$-directional distance is reduced by $\sim 0.2 Y$ for $\vrot\sim 150 \ \ekms$, where the factor arising from the second term inside the square root of figure \ref{eqY} is negligible in the present area.
A similar correction is necessary, but in the opposite sense, if we adopt a different rotation velocity, so that higher $\vrot$ results in larger $Y$ distance.
Thus, if we correct for the $\vlsr$ displacement of $\delta v \sim -30 \ \ekms$ in the present data, the $Y$ directional extent of the Sgr B bubble may be reduced by a factor of $Q\sim 1+\delta v/\vrot \sim  0.8$.

\section{Formation mechanisms: Molecular bow shock, cometary HII region, and oval shock wave}
\label{sec4} \label{models}

\begin{figure*}   
\begin{center}    
(A)\includegraphics[width=7.5cm]{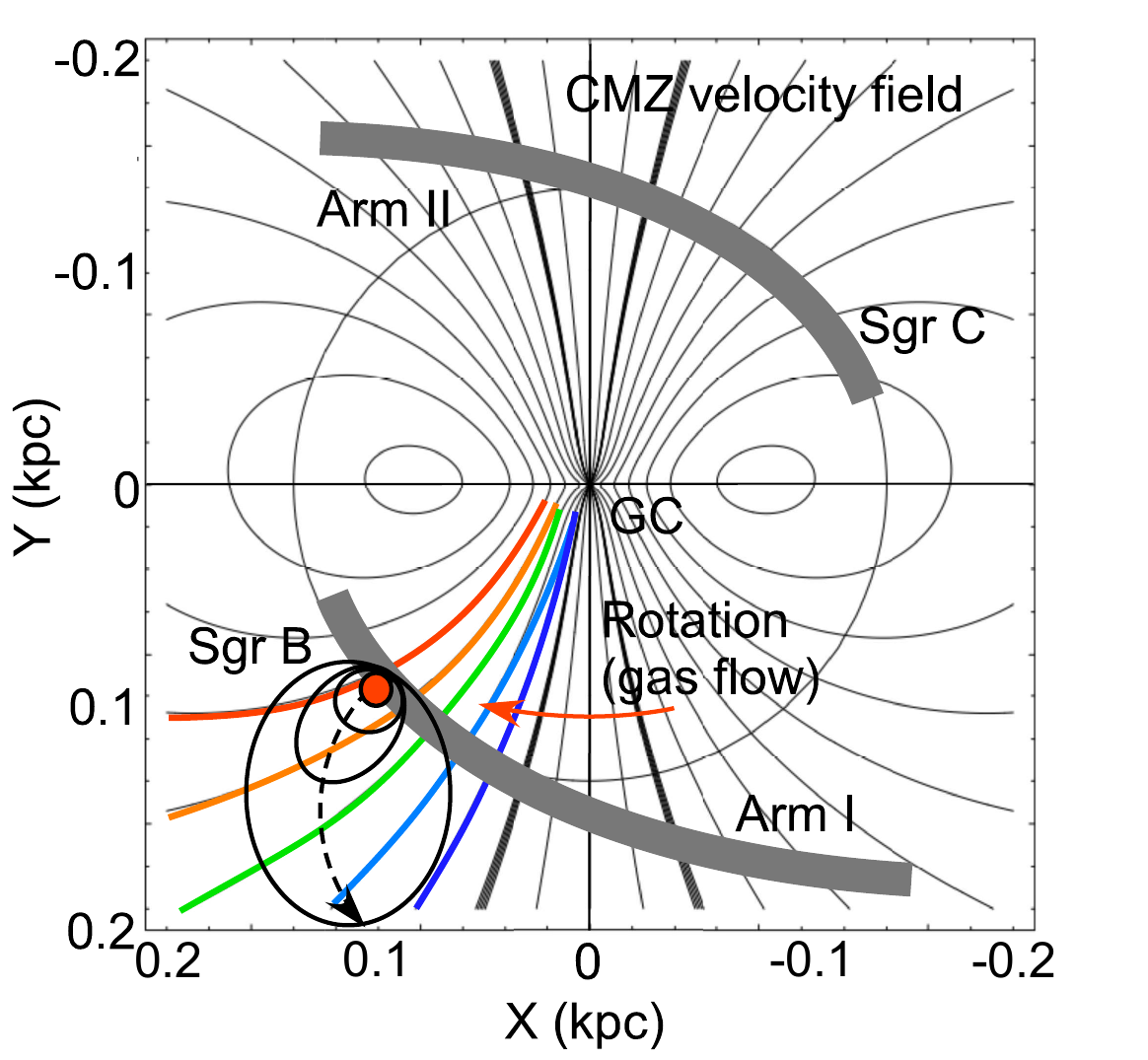} 
\hskip 5mm
(B)\includegraphics[width=7.4cm]{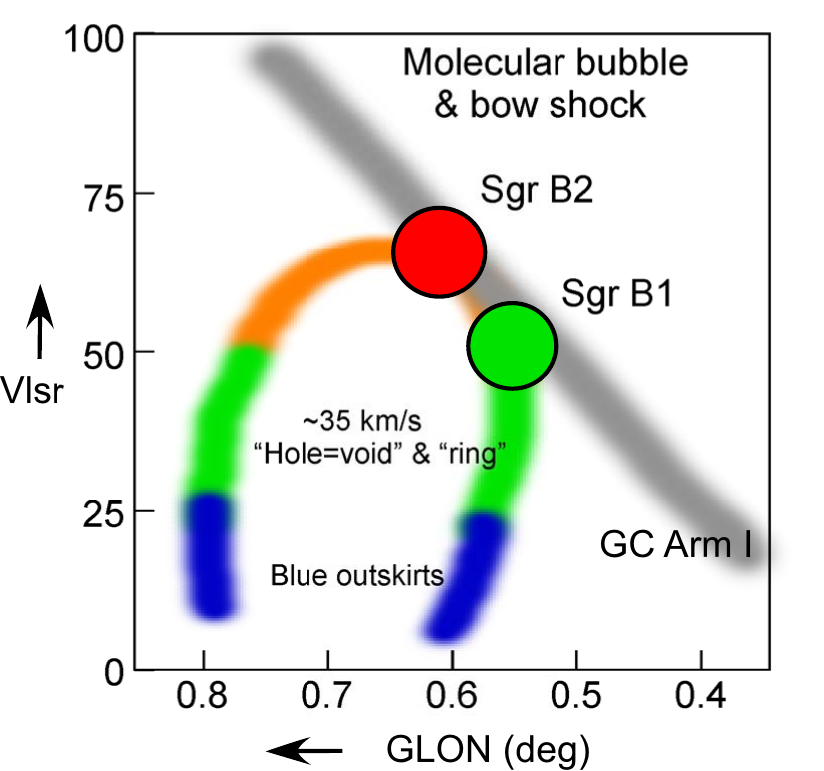}   \\
\vskip 5mm
(C)\hskip 5mm\includegraphics[width=7.5cm]{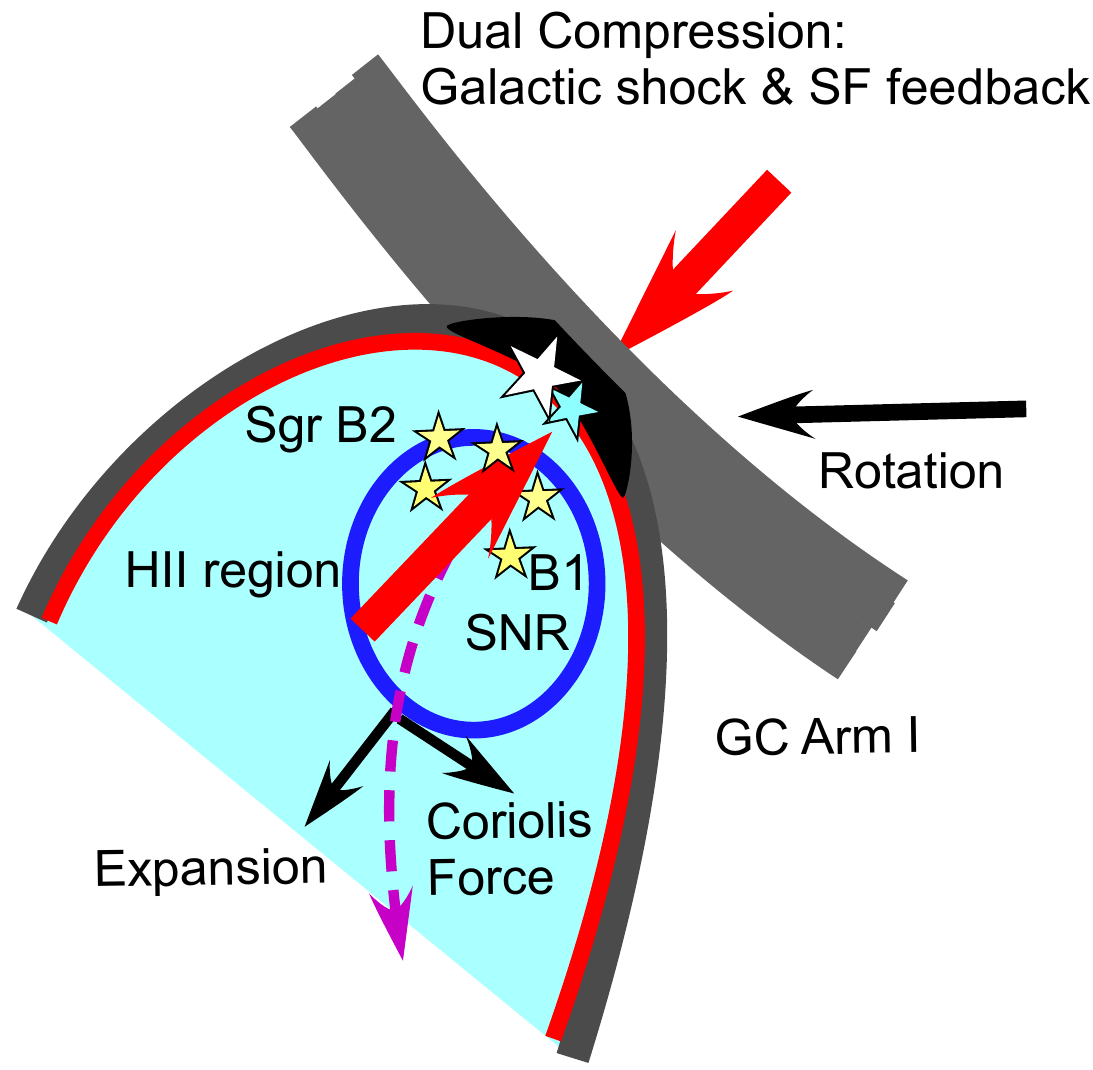} 
(D)\includegraphics[width=7.5cm]{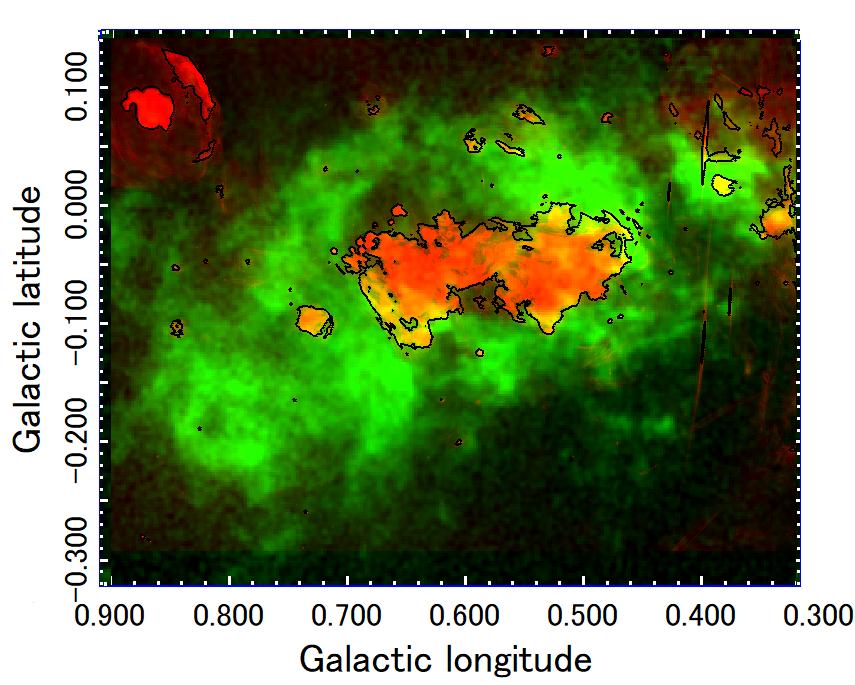} \\
\vskip -3.5cm
-----------------------------------------------------------~~~~~~~~~~~~~~~~~~~~~~~~~~~~~~~~~~~~~~~~~~~~~~~~~~~~~~~~~~~~~~~~~~~~~~~~~~~~~~~~~~~~~~~~~~~~~~~~~~~~~~~~~~~~~~~~~~~~~~~~~~~~~~~~~~~~~~~~~~~~~~~~~~~~\\
\vskip 3.5cm

\end{center}
\caption{(A) Schematic velocity field and GC Arms I and II, and Sgr B projected on the galactic plane.
(B) Schematic longitude-velocity diagram of Sgr B2, B2, Sgr B molecular complex. Red color represents $\vlsr \sim 60$--70 \kms, green $\sim 30$--50 \kms, and blue $\sim 20$--30 \kms. The grey line is the GC Arm I on which Sgr B1 and B2 are located. The $\sim 35$-\kms "hole and ring" \citep{bal88} is here interpreted as the slice of the bubble/cylinder at the velocity.
(C) Dual-side compression (DSC) of  molecular cloud by the Galactic shock wave and stellar feedback. Red arrow from right top represent compression perpendicular velocity $(\vrot-\vpat) \sin \ p\sim 30$ \kms, red arrow from left bottom.  
\red{(D) MeerKat continuum map in red with black contours at 1 mJy beam$^{-1}$ overlaid on \coth 35 \kms channel map in green (as figure \ref{fig-rgb} bottom left). The -molecular cloud in green approximately represents cross section of the bow-shock cone along the straight line in panel (C), while the continuum emission is the integration along the line of sights. } }
\label{fig-illust}	 
\end{figure*}

We have shown that the Sgr-B molecular complex has a conical-horn structure of molecular gas open toward the Sun from the Sgr-B2 cloud in the farthest end on the line of sight. 
In this section we consider possible formation mechanisms of such an open bubble structure, although their local velocities are smaller than that of the Galactic rotation.
We here discuss the following three cases:  
(a) molecular bow shock  (MBS) \citep{sof18}, 
(b) giant cometary-HII region (GCH)
\citep{1985PASJ...37..507S,sof18,2019PASJ...71..104S,2019PASJ...71S...1S}, 
and
(c) oval shock wave \citep{sof90,sof23g18}. 
In the models, we assume that the complex has expanded around Sgr-B2 cloud in the flow on the Galactic rotation with the CMZ, which encounters the Galactic shock wave along the GC Arm I. 
We here also consider an expansion due to the thermal pressure of ionized hydrogen gas (HII region) heated by the central OB stars, and assume it to be on the order of the sound velocity of 
$v_{\rm bub}\sim 10\ekms$.
Therefore, the bubble's age is 
$t_{\rm bub} \sim r/v_{\rm bub}\sim 2$ My for
$r_{\rm bub}\sim 20$ pc.
This velocity is smaller than the rotation velocity, $\vrot \sim 150 \ekms$. 
If we take into account the expansion and differential rotation simultaneously, the front shape will be deformed in the sense that the outskirts is trailed counter-clock wise seen from the north Galactic pole due to the Coriolis force on the order of $v_{\rm bub} \Omega t_{\rm bub}^2 \sim 50 \epc$  for $\vrot\sim 150 \ekms$ and radius $\sim 120$ pc \citep{sof13}, where $\Omega$ is the angular velocity of the rotation. 
Therefore, the deformation due to Galactic rotation is on the same order of the bubble's size.
Figure \ref{fig-illust} illustrates the model and mechanism for the formation of the Sgr B complex proposed in this paper based on the analysis in the previous section.
 
Astrophysical bow shock is a classical phenomenon in objects interacting with supersonic gas flow in the interstellar space
\citep{1975Ap&SS..35..299D,1990ApJ...353..570V,1995Ap&SS.224..151O,wil96,2002ApJ...575..928A,2002AJ....123..362R}.
On the scale of star-forming site, it is observed as cometary HII regions tailing down-stream direction from the OB cluster on the head 
\citep{2000ApJ...533..911F,2006ApJS..165..283A,2015A&A...582A...1D,zhu15,2017MNRAS.466.4573S}. 
On galactic scale, large bow shocks at the spiral arms are observed in the tangential direction of the 4 kpc arm as associated with the SF region W43 ($l\sim 30\deg$) and the Norma arm ($l\sim 22\deg$)
\citep{1985PASJ...37..507S,2019PASJ...71S...1S,2019PASJ...71..104S}. 
MBS are also observed in spiral galaxies such as M83, M51, and M33, and are ubiquitous in galactic shock-wave arms \citep{sof18}.
Recent far-infrared images with the JWST of spiral galaxies such as M74 (NGC 628) exhibit numerous MBS and similar structures  \citep{2023ApJ...944L..17L,2023ApJ...944L..22B}. 
An MBS is a concave arc or conical cylinder of molecular gas around an expanding HII region formed in the down-stream side with respect to the galactic shock waves.  
In addition to these individual phenomena, we further consider that the galactic shock-wave theory \citep{fuj68,rob69} applies to the arm-size structures, which include GC Arm I and II. 
 
\begin{figure} 
\begin{center}   
\includegraphics[width=8.5cm]{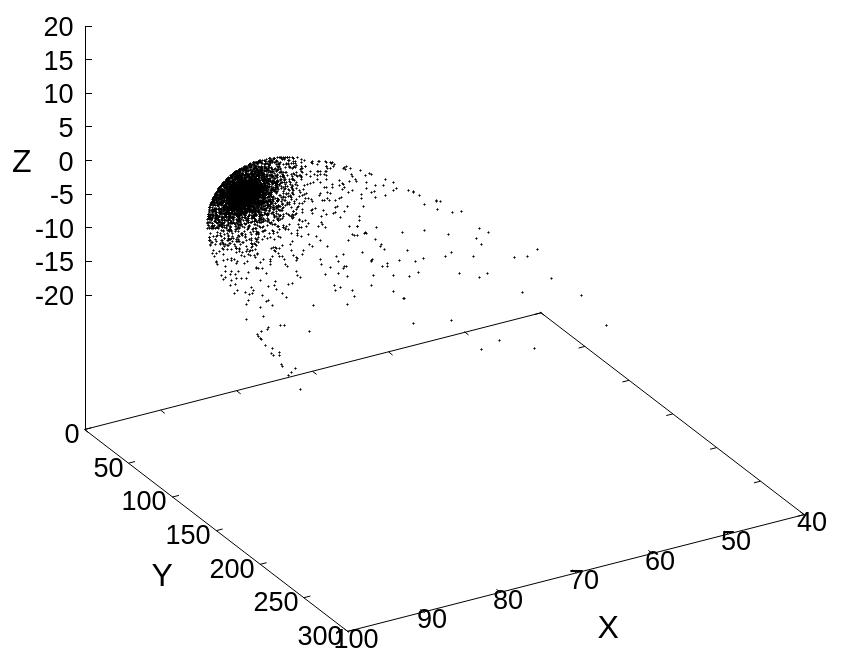} \\
\vskip 2mm
\includegraphics[width=8.5cm]{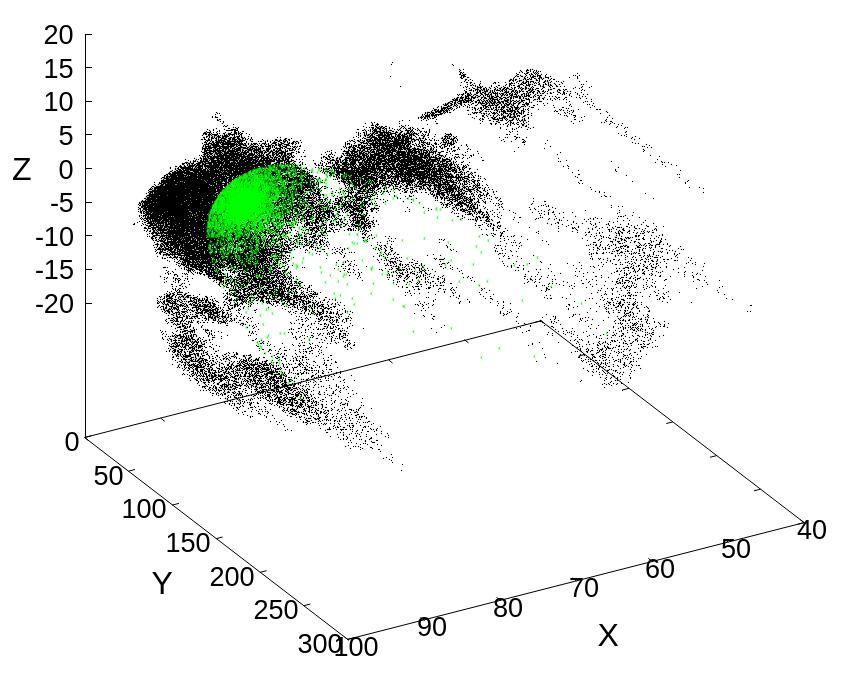}   
 \end{center}
\caption{[Left] Bow-shock model.
[Right] Overlay of 3D Sgr B molecular complex and bow-shock model.  }
\label{fig-3d-model}
	\end{figure}     
          
\subsection{Molecular bow shock (MBS) in a supersonic flow}
\label{sec-wilkin}

In our paper \citep{sof18} we modeled the MBS G30.5 by applying the \citet{wil96}'s 
analytical model for stellar-wind bow shock.
The distance $Q$ of a bow front from the wind source is related to the elevation angle $\phi$ through
\def\Rbow{R_{\rm bow}}
\be
Q(\phi)=\Rbow\ {\rm cosec}\ \phi \ \sqrt{3(1-\phi\ {\rm cot}\ \phi}).
\label{eq-Q}
\ee 
Here, $\Rbow$ is the stand-off radius defined as the distance of the front on the galactic plane from the wind source, which is measured as the smallest curvature of the bow head facing the gas flow. It is related to the momentum injection rate, or the mass loss rate of the central stars through \citep{vin22} 
\be
\dot{M} = 4 \pi \Rbow^2 v_{\rm CMZ}^2 \rho_{\rm CMZ}/v_{\rm w},
\ee 
where $M$ is the mass of the central stars, $\Rbow$ the bow-radius at the head, $v_{\rm CMZ}$ is the inflow velocity of the CMZ gas, $\rho_{\rm CMZ}$ its density, and $v_{\rm w}$ is the wind velocity.

\def\Vpat{V{\rm pat}} \def\Vrot{V_{\rm rot}}

\begin{figure} 
\begin{center}    
\includegraphics[width=7.5cm]{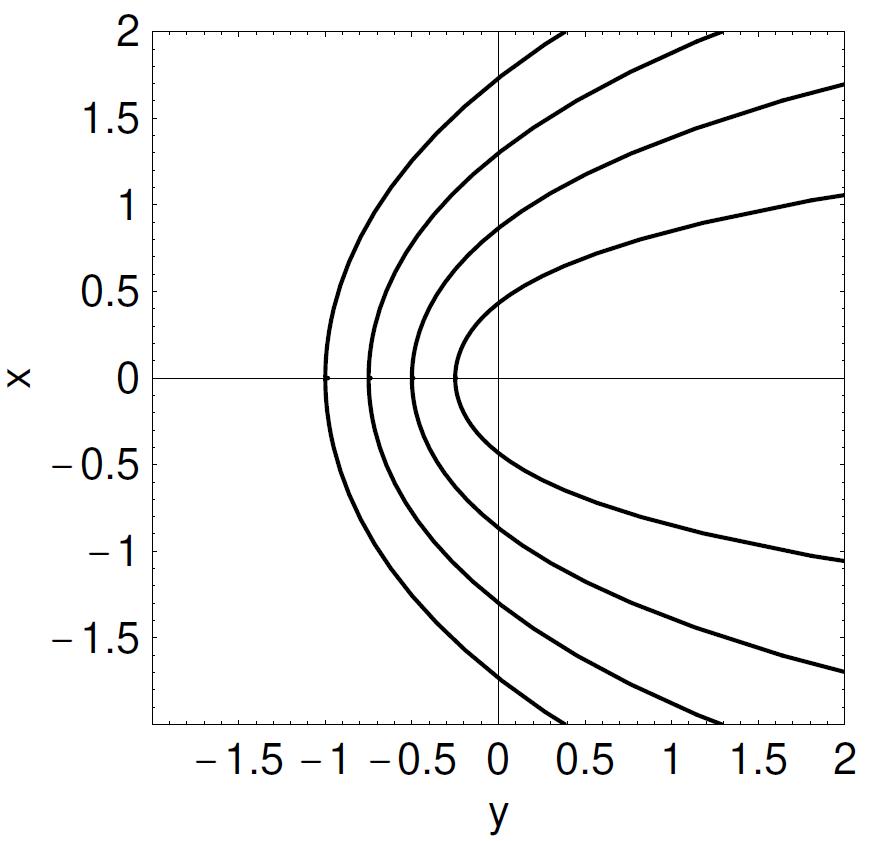}  
\includegraphics[width=7.5cm]{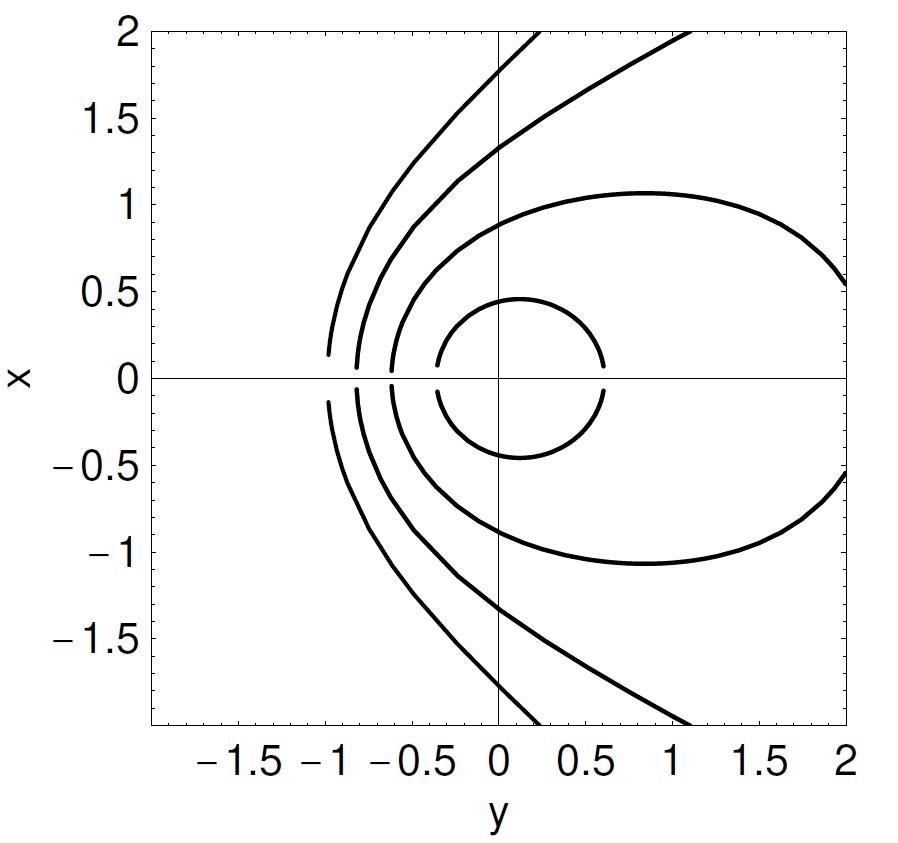}   
 \end{center}
\caption{[Top] Bow-shock front by supersonic flow encountering with stellar wind given by equation \ref{eq-Q}, where $x$ and $y$ are measured from the star in unit of the stand-off radius $R_{\rm bow}$ \citep{wil96}. 
[Bottom] Expanding HII-region front in a density gradient proposed for the molecular bow shock and cometary HII regions \citep{sof18}. The scale is arbitrary. 
}
\label{fig-frontbow}
	\end{figure}

\begin{figure} 
\begin{center}     
\includegraphics[width=7.5cm]{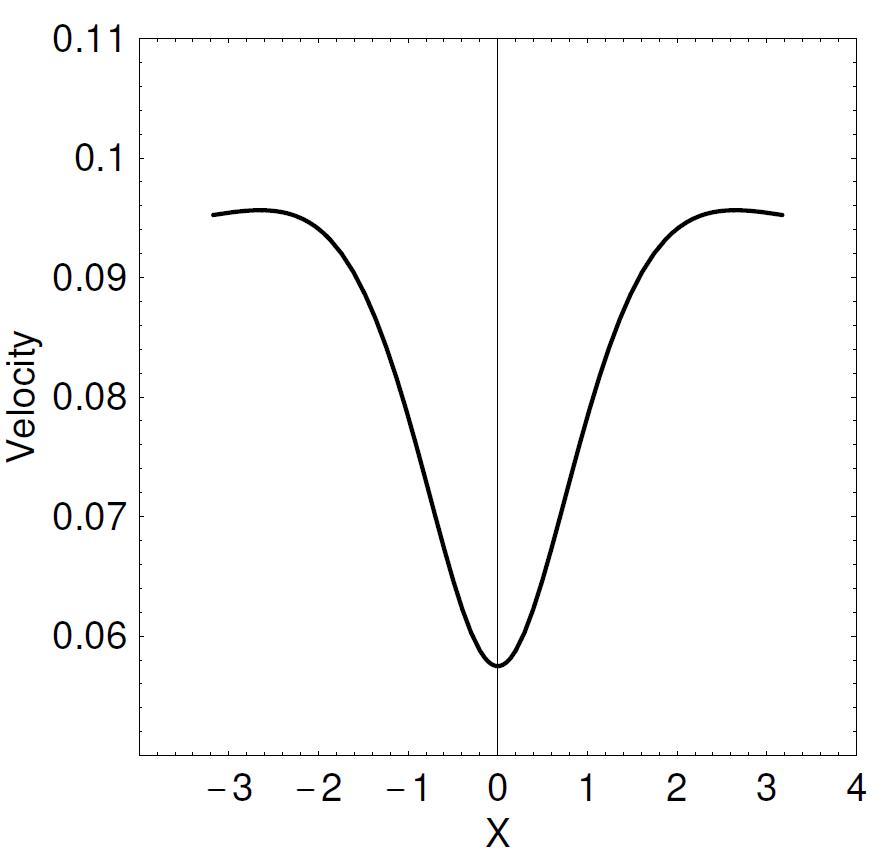}
\includegraphics[width=7.5cm]{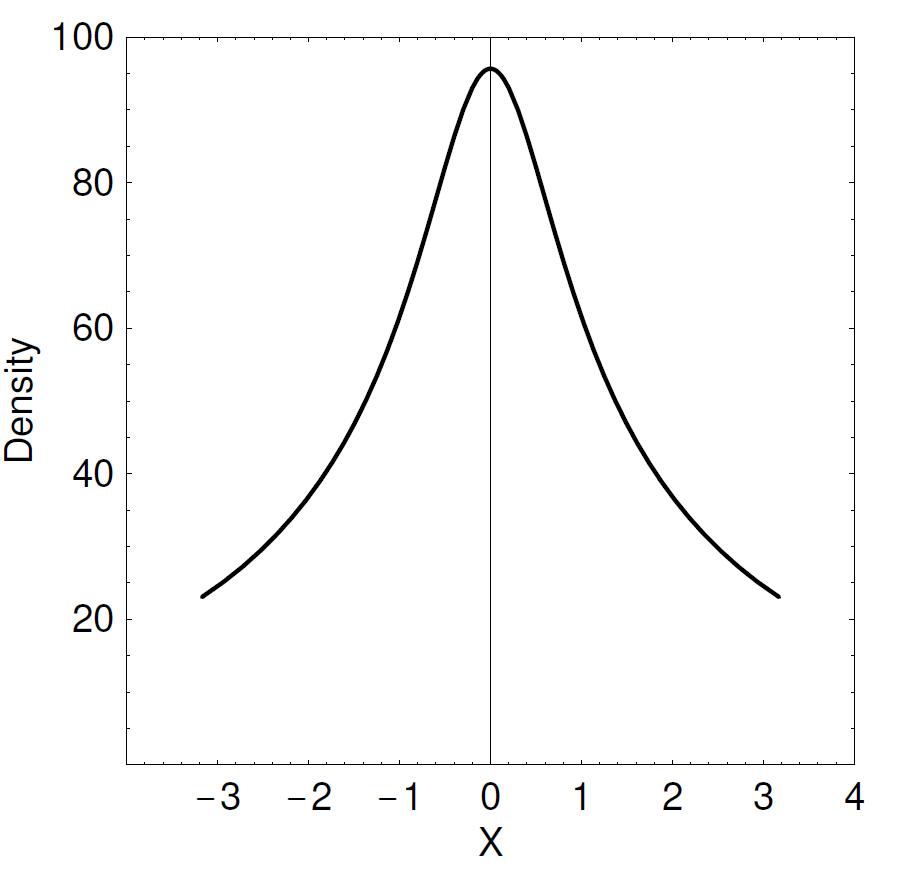} 
 \end{center}
\caption{[Top] Tangential velocity of the flow along the bow-shock plotted against the distance from the bow axis corresponding to figure \ref{fig-frontbow} following \citet{wil96}. Vertical axis is arbitrary.
[Middle] Corresponding surface-gas density against. Vertical axis scale is arbitrary.
} 
\label{fig-bowveloden}
	\end{figure}

Figures  \ref{fig-3d-model} and \ref{fig-frontbow} show the bow's shape, where the wind parameter is determined to be $\Rbow\sim 1.25$ pc in order to yield a reasonable fit to the observed bow of Sgr B as in figure \ref{fig-3d-model}. 
The top panel of figure \ref{fig-bowveloden} shows variation of the tangential velocity as a function of the distance from the bow axis, which approximates the position-velocity diagram seen from the bow axis.
The "V" shape observed in the position-velocity diagrams (figures \ref{fig-lvd}) is well reproduced.
The bottom panel shows the variation of the density on the bow surface.
We here assumed the background gas density
$\rho_{\rm CMZ}\sim 100 {\rm H_2 \ cm^{-3}} \sim 4.6\times 10^{-22}{\rm g\ cm^{-3}} $,
wind velocity $v_{\rm w}\sim 2000$ \kms,
galactic-shock velocity of 
$v_{\rm CMZ}\sim (\Vrot - \Vpat)\sin \ p \sim 17 \ekms$
with $\Vrot\sim 150 \ekms$, $\Vpat\sim 1/2 \Vrot$ and $p\sim 12\deg$ being the rotation velocity of the CMZ, pattern speed, and pitch angle of the spiral arm, respectively. 
Thus, we obtain $\dot{M}\sim 2.0\times 10^{-5} \Msun {\rm y}^{-1}$.
A couple of O stars and continuously-flowing molecular gas in the CMZ drive the bow structure of Sgr B cloud.

\subsection{Giant cometary HII region (GCH)} 

Conical structure can be modeled also by a deformed Str\"omgren sphere in a medium with density gradient \citep{sof18}, which yields a qualitatively similar result to full hydrodynamical simulations \citep{zhu15}.
The Str\"omgren radius in a uniform ISM is related to the luminosity of the central OB stars by 
\def\nuv{\mathcal{N}_{\rm UV}}
\be
\nuv \simeq \frac{4 \pi}{3} \Rhii^3  \ni\ne\ar ,
 \label{strsph}
 \ee 
where $\nuv$ (s$^{-1}$) is the UV photon number radiated by the OB stars per time, $\ar \sim 4\times10^{-13}  {\rm cm^{-3}s^{-1}}$ is the recombination rate, $\ni$ and $\ne$ are the ion and electron densities, respectively.

We here assume that the equation holds in each small solid angle at any direction in which the density is assumed to be constant. 
By neglecting dynamical motion of gas, this approximation gives a qualitative shape of the expanding HII-region front. 
  
Assuming that the ion and electron densities are related to the neutral gas density $n$ through
$
\ne\sim \ni \sim n ({T_{\rm n} /T_{\rm e}}),
\label{ne}
$
where $T_{\rm n}$ and $T_{\rm e}$ are the temperatures of neutral and HII gas, we define the representative radius $\Rhii$ as the equilibrium radius of a spherical HII region in uniform gas with density $n$ by
\be 
\Rhii \simeq \left[{3 \nuv \over 4 \pi \ar n^2 } \left(\Te /\Tn \right)^2\right]^{1/3}.
\label{Rhiineu}
\ee
Rewriting $\nuv \sim \L/h\nu$ with $\L$ and $h\nu$ being the luminosity of the central OB stars and UV photon energy over 912 A, respectively, 
we have
\begin{equation}
\Rhii \sim 96 
\(\frac{\L}{10^4 \Lsun} \)^{\frac{1}{3}} 
\( \frac{n}{10^2{\rm H \ cm^{-3}}} \) ^{-{\frac{2}{3}}}   
\(\frac{\Tn}{20\ K}\)^{-{\frac{2}{3}}} 
\(\frac{\Te}{10^4\ K}\)^{\frac{2}{3}}
\red{(\rm pc)}.  
\label{Rhiipc}
\end{equation}
The front shape varies according to the parameter $\Rhii \propto L^{1/3}n^{-2/3}$, and attains an elliptical shape elongated in the lower-density direction \citep{sof18}.        
On the other hand, given the gas density, $n\sim 10^3$ \htwo cm$^{-3}$, the UV luminosity $\L$ of exciting OB stars can be  estimated for the measured $\Rhii\sim 20$ pc as 
$\L  \sim 10^4 \Lsun$.
The bottom panel of figure \ref{fig-frontbow} shows an example of such deformed HII front proposed for the cometary HII regions in M83 \citep{sof18}.
 
\subsection{Oval shock front driven by HII region and/or SNR}

The third mechanism is a shock wave driven by supersonic expansion of the HII region and/or supernova remnant in Sgr B2 and B1.
In order to examine this model, we apply the Sakashita's method in order to trace the propagation of an adiabatic-shock wave in an in-homogeneous medium originally developed to trace the propagation of a shock-wave from a galactic nucleus through the disc \citep{sak71}.
The method has been applied to various galactic objects including an interstellar shock wave from a molecular cloud.
We mimic here the original dense cloud by an elongated density concentration in a uniform medium as investigated in our recent paper \citep{sof23g18}.

\begin{figure*}
\begin{center}     
\includegraphics[width=8cm]{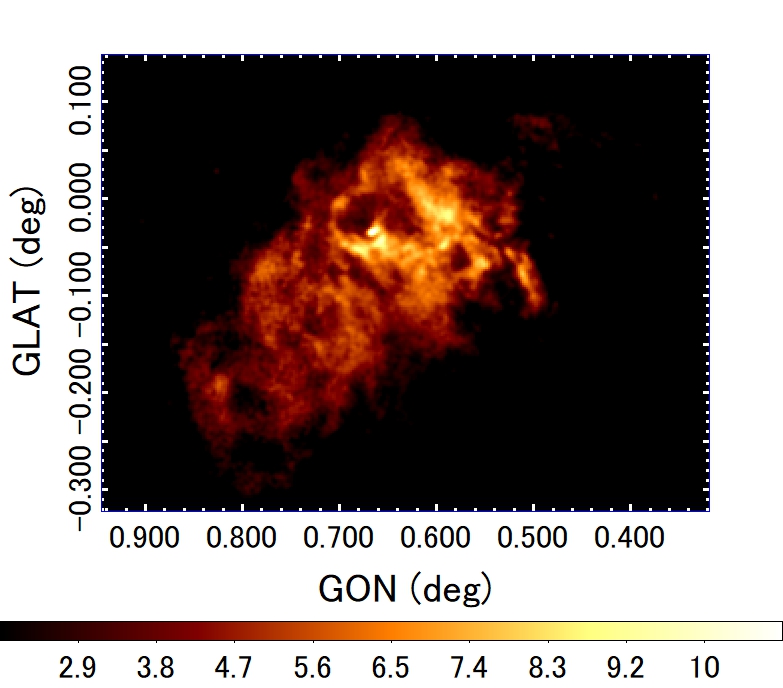} 
\includegraphics[width=8cm]{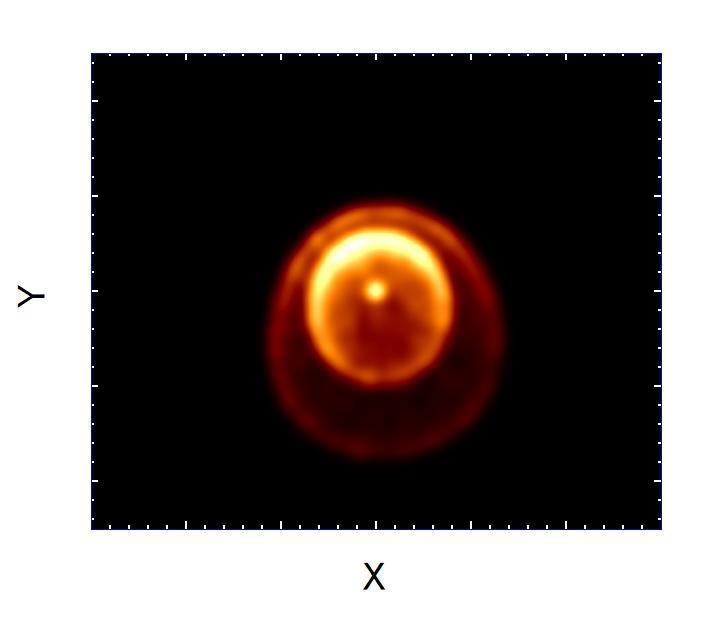}  \\  
\vskip 3mm
\includegraphics[width=7.3cm]{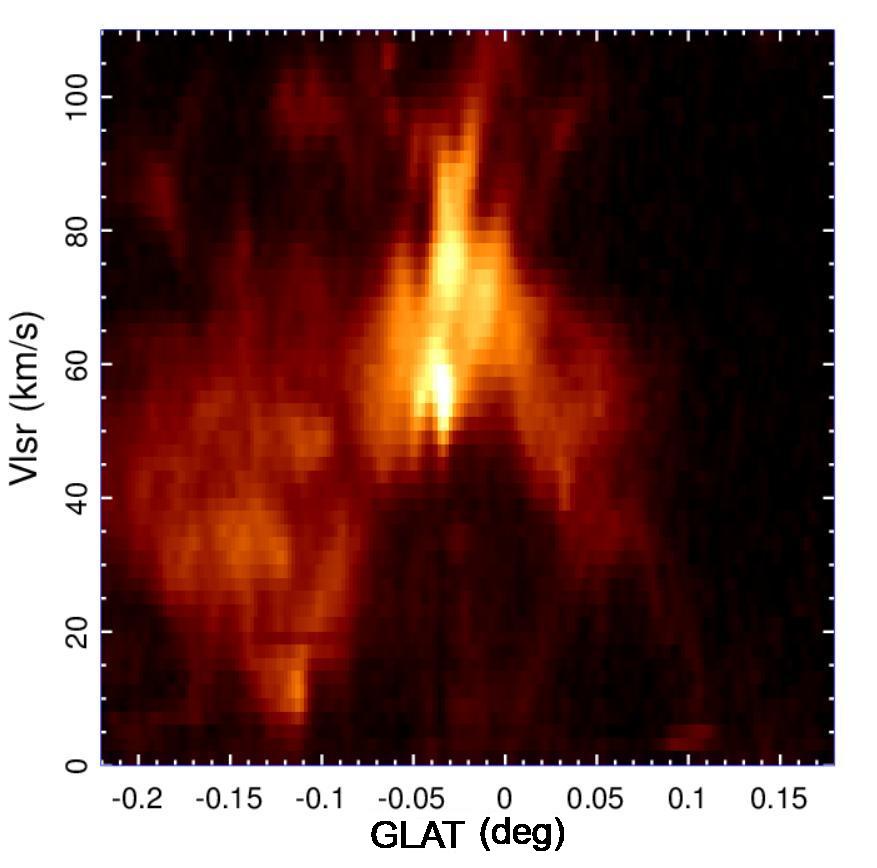}  
\hskip 4mm
\includegraphics[width=7cm]{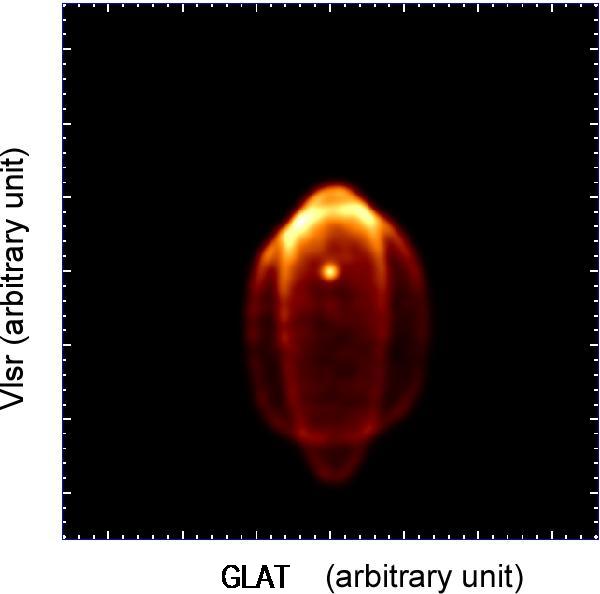}  
\end{center}
\caption{
[Top left] Sgr B molecular complex in the \coth-line emission. 
[Top right]
\red{Calculated shock-wave fronts projected on the $(X,Y)$ plane
due to an explosion at the cloud's edge.
Fronts at two epochs at $t=1$ and 2 time units from the explosion are presented in order to show the evolution.
The shell expands toward the low-density side (toward bottom), attaining a lopsided open cone, and mimics the molecular bubble around Sgr B.
Scales are arbitrary and the time is normalized by $r_0/v_0$ with $r_0$ and $v_0$ being the unit radius and expansion velocity in a uniform medium with unit density, respectively.
[Bottom left] Observed latitude-velocity diagram across Sgr B2.
[Bottom right] Calculated position-velocity diagrams of the above shock fronts seen from the opening end of the cone.  
This model demonstrates that the bubble and V-shaped PVD in Sgr B can be produced also by such a lopsided explosion, which suggests that the real Sgr B may be a hybrid of asymmetric bubbles formed by various mechanisms.}
}
\label{fig-shock}	
\end{figure*}

In the present model, energy on the order of $\sim 10^{50}$ erg is released at a point on the oblique edge of the cloud, and the propagation of the shock front is traced using Sakashita's method.
Figure \ref{fig-shock} shows an example of such oblique shock front projected on the sky and corresponding position-velocity diagram (right panels) as compared with the intensity distribution and latitude-velocity diagram of the Sgr-B complex (top left and bottom right, respectively).

The shock front expands spherically in the initial stage, but after a while it is deformed to attain an oval structure, and expands faster in the direction to decreasing density.
The top-right panel of figure \ref{fig-shock} shows such elongated oval projected on the sky, while the projected shape is nearly spherical because the line of sight is nearly parallel the oval's major axis in this model.

The bottom-right panel shows a position-velocity diagram across this front as seen from the major axis (observer), and is compared with the observation (left panel).
The shock front draws a lopsided oval, mimicking the observed LVD.

We have thus shown that an explosive event at the cloud edge can explain the lopsided bubble structure as well as the position-velocity behavior.
In these calculations, however, we did not take into account the rotation of the ambient medium, which would deform the bubble shapes on the sky as well as in the LVD.
Although the present models may explain the observed shapes qualitatively, full dynamical simulations are necessary to quantitatively reproduce the observed properties.

\section{Discussion}
 \label{sec5}
 
\begin{figure*}   
\begin{center}  
(A) \includegraphics[height=6.5cm]{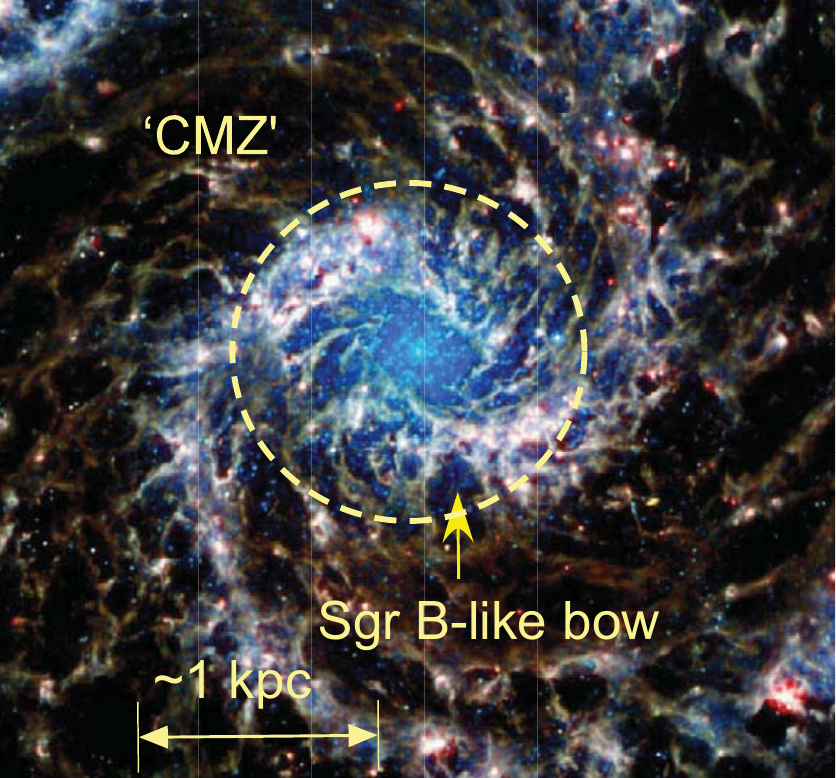}   \hskip 5mm
(B) \includegraphics[height=6.5cm]{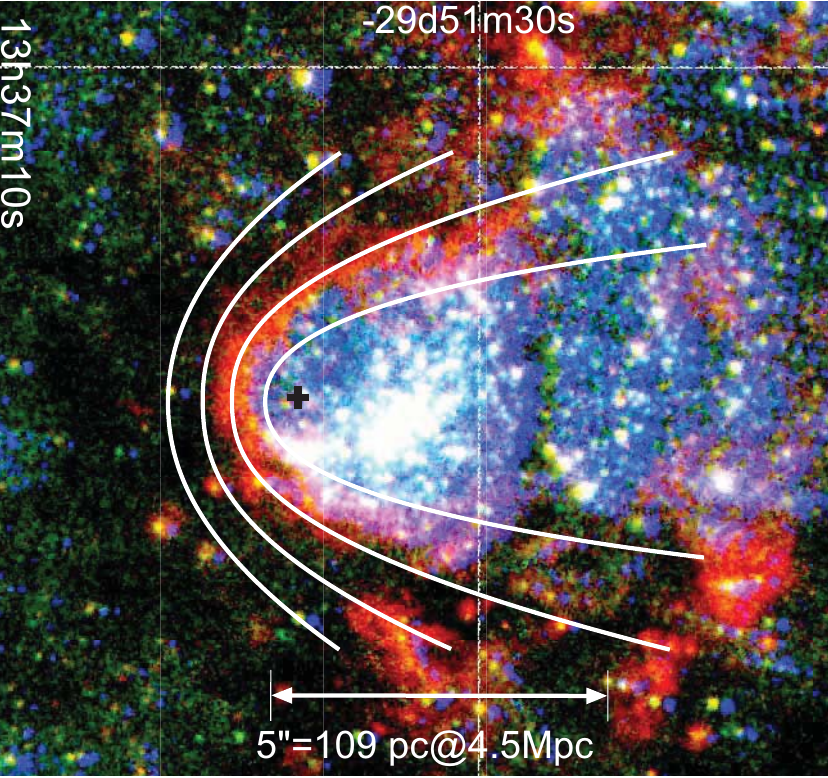} \\
\vskip 5mm \hskip -7mm
(C)\includegraphics[height=6.7cm]{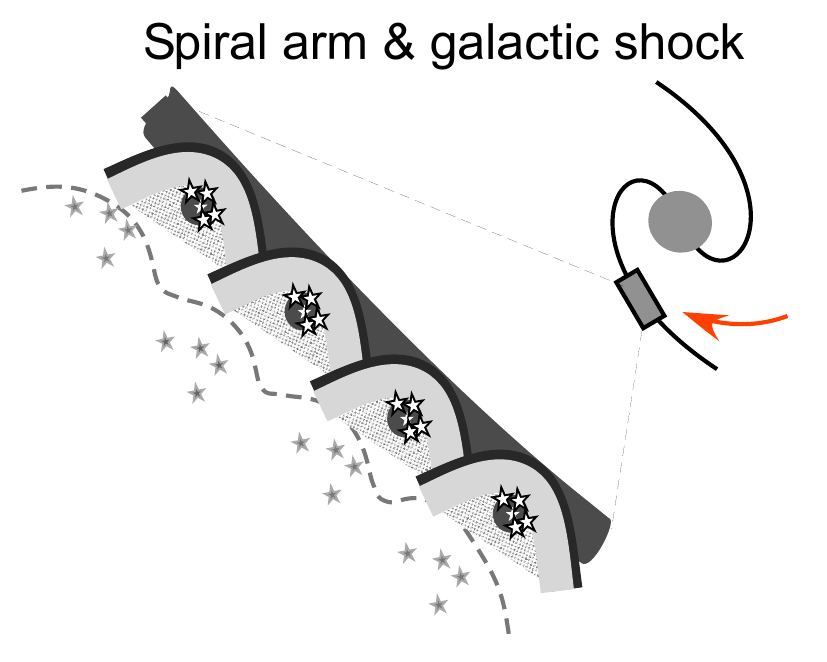}  
(D) \includegraphics[height=6.7cm]{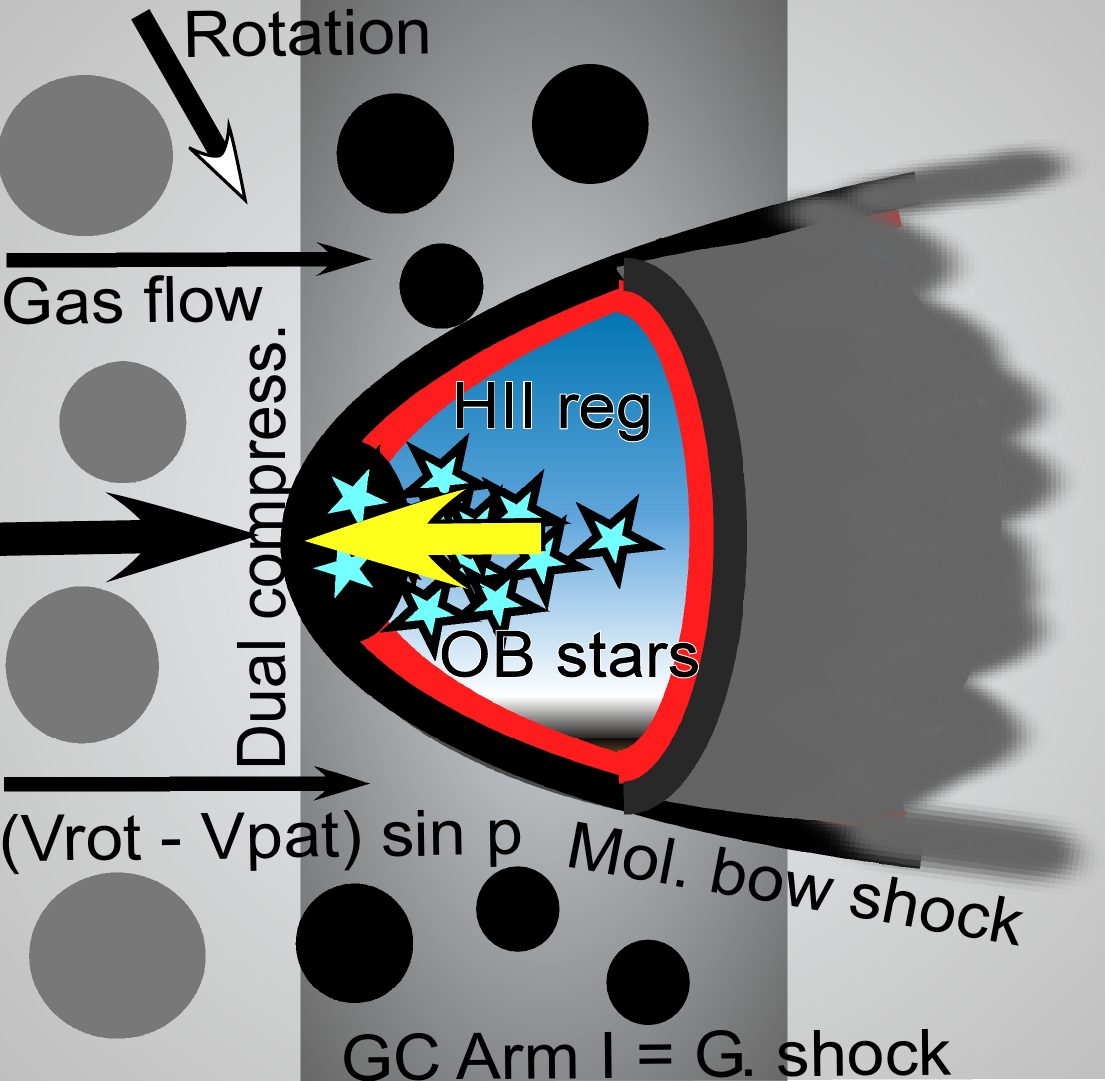}  
\end{center}
\caption{
(A) Molecular bow shock similar to Sgr-B molecular complex in the mid-infrared image of the spiral galaxy M74 (NGC 628) taken with the JWST \citep{2023ApJ...944L..17L,2023ApJ...944L..22B}. Numerous MBS are expanding toward the down-stream sides of the spiral arms. 
(B) Giant cometary HII region (GCH) sheathed by an MBS in the optical image of M83 with the HST \citep{2017AJ....154...51M}.
The front is fitted by a bow-shock as indicated by the white lines \citep{sof18}.
\red{
(C) Illustration of a chain of MBS+GCH along a spiral galactic shock wave.
(D) Illustration of the MBS+GCH model for Sgr-B molecular complex. 
Dual-side compression (DSC) of the stagnated molecular cloud triggers the star formation.
Velocity vectors (arrows) are relative to the rotating arm, where $V_{\rm rot}$, $V_{\rm pat}$ and $p$ are the rotation velocity, pattern speed and arm's pitch angle, respectively.}}
\label{fig-msb-m74m83}	
\end{figure*}

As described in section \ref{intro} there have been three models so far proposed for explaining the "hole" and "ring" of the Sgr B molecular complex, which are
(i) the local perturbation model \citep{bal88},
(ii) expanding-bubble and stellar feedback model \citep{sof90}, and 
(iii) cloud-collision model \citep{has94,sat00,eno22}.  
The purpose of this paper was to explore the bubble/shell model (ii) in the light of recent CO-line and radio-continuum data.
We explained the complex by an MBS (molecular bow shock) induced by the galactic shock wave in the CMZ, which produced the a horn-like bubble structure assisted by the energy feedback of star formation in Sgr B1 and B2 associated with a GCH (giant cometary HII region). 
We have shown that the model can explain various properties observed in the complex, which we summarize below.
We summarize the present model in analogy to the bubbles and bow-shock structures as observed in M74 and M83 in figure \ref{fig-msb-m74m83}.

\subsection{Galactic shock waves, fins, and voids}

The galactic shock wave induced by the density waves of spiral arms  \citep{fuj68,rob69,rob79,wad11,baba17} and bar potential  \citep{soren76,sorma15} are the well-studied and  established phenomenon in the gas dynamics of galactic discs, which also applies to the Galactic centre \citep{sorma20}. 
High-resolution numerical simulations show that fin structures are bifurcated from the spiral shock waves and compose void features \citep{baba17}.
Recent far infrared imaging with the JWST of nearby spiral galaxies revealed numerous "voids" in the dusty spiral arms, which are generally curved toward the up-stream direction \citep{2023ApJ...944L..17L,2023ApJ...944L..22B}. 
We point out that the voids are far-infrared views of the MBS (molecular bow shocks) sheathing GCH (giant cometary HII regions) \citep{sof18}.
We argue that the MBS+GCH structure is ubiquitous also in the Milky Way \citep{1985PASJ...37..507S,2019PASJ...71S...1S} and the CMZ as observed as Sgr B complex here. 
The result of the present paper lies on the extension of these general views of the molecular gas distribution in spiral arms and central bar-shocked regions of galaxies.
 
\subsection{"V"-shaped position-velocity diagrams around stagnation}
\label{sec-vshape}

The linear alignment of the Sgr B1 and B2 clouds on the LVD (figure \ref{fig-lvd}) indicates that the two clouds are located on the GC Arm I and are rotating circularly in the Galactic disc. 
It is remarked that a number of compact HII regions are associated with Sgr B1 and B2 as observed in radio recombination lines, which lie on the same LV ridge \citep{meh92,meh93}.
This means that Sgr-B2 cloud composes a part of the main structure of Arm I in regular Galactic rotation, but it is not a flying-by object, as the collision model postulates.

The "V"-, or "$\Lambda$", shaped LVD near Sgr B2 (figure \ref{fig-lvd}) is explained by intersection of the main ridge of GC Arm I and the horn-like bubble of the Sgr B complex, where the main LV ridge (figure \ref{fig-lvd})  nests the HII regions Sgr B2 and B1.
The HII regions are rotating on the Arm, exhibiting the LSR-velocity gradient $d\vlsr/dl\simeq 150 \ekms/{\rm deg}\simeq 11 \ekms\ \epc^{-1}$.
The western wing of the $\Lambda$ is, therefore, the main ridge of the rotating Arm, but not due to local disturbance from outside.
The molecular bubble shows up as a bow-shaped arc in the LVD in touch with the GC Arm main ridge (figures \ref{fig-lvd} and \ref{fig-illust}), where the western wing composes a fainter $\Lambda$, bifurcating from the main Arm toward decreasing velocity.
The LV behavior is thus naturally explained by the galactic rotation and extended bubble in the rotating disc.

The latitude-velocity diagrams (BVD) across Sgr B2 also exhibit V shape as shown in figures \ref{fig-lvchan}, \ref{fig-lvd} and \ref{fig-shock}.
It is interesting to point out that the width of the "$\Lambda$" in the BVD is $\sim 0\deg.15$ (20 pc), about a half of the width in LVD of $\sim 0\deg.3$ (40 pc).
This flattened conical shape in the $(l,b,\vlsr)$ space can be naturally understood as due to a bow shock by the galactic shock wave flattened along the disc plane.

\red{We mention that the V-shaped bow structure is also clearly seen in the HCN $(J=4-3)$ line, which is the highest-density tracer among the lines ever used for the CMZ \citep{2018ApJS..236...40T}, as shown in panel (D) of figure \ref{fig-lvd}.
Moreover, the stagnation point (cap) is saturated in the line, even producing deep absorption against the continuum emission of Sgr B2.
A similar bow feature with absorption at Sgr B2 is also recognized in the CS line \citep{1999ApJS..120....1T}.
The fact indicates that the gas density attains a steep maximum at the stagnation point associated with the active star formation of Sgr B2, where the critical density is $\gtrsim 10^7$ H$_2$ cm$^{-3}$, while the effective density would be one or two orders of magnitude lower due to the photon trapping effect \citep{2018ApJS..236...40T}. } 

All the models for Sgr B complex presented in section \ref{models} indicate similar behaviors of flow speed in the position-velocity diagram. 
In the bow-shock model, the supersonic flow is stagnated at the bow head, where the velocity is zero with respect to the shock wave, and increases along the bow surface with the distance from the bow axis \citep{wil96}.
As shown in figure \ref{fig-bowveloden}, the theoretical calculation of the tangent velocity along the bow as a function of the distance from the bow axis well explains the observed V shape in figure \ref{fig-lvd}. 
We also point out that the theoretical density variation along the bow, as shown in the bottom panel of figure \ref{fig-bowveloden}, is well observed in the BVD across Sgr B2 (figure \ref{fig-lvd}).
In the expansion models due to HII regions, a similar V shape appears according to the lopsided expansion at different velocities depending on the directions from the centre (figure \ref{fig-shock}).

Thus, in general, a V-shaped PVD is associated with stagnated flow of gas, which is observed in the galactic shock waves, in bow shocks around a star, around an HII region, around a cloud, and in a cloud under collision with another cloud.
Lopsided expanding shells around HII regions in non-uniform medium also exhibit partially elliptical PVD, mimicking V shape.
Therefore, the V-shaped PVD is a general astrophysical phenomenon in the ISM.
We mention that a similar property is observed in the cloud-collision, where the flow pattern around the colliding body is identical to that in general formed around the bow shock.

\subsection{Complicated velocity dispersion in the Sgr-B2 cloud}

The moment 2 map in figure \ref{fig-moment} indicates systematically lower velocity dispersion at $\sigma_v\sim 10$ toward the Sgr-B2 cloud 
\red{ over $0\deg.07$ (10 pc)} diameter than the surrounding cloud with $\sigma_v\sim 18$ \kms. 
This fact is opposite to that expected for a cloud collision which induces stronger turbulence and velocity dispersion between the two clouds.
This fact is in favor of the expanding-bubble model, where the central high-density Sgr-B2 cloud composes the major structure of the GC Arm I, and the expelled clouds from the Arm compose the expanding bubble with higher velocity dispersion.

On the other hand, the very central clump of diameter of $0\deg.02$ (3 pc) at $(l.b)=(0\deg.66,-0\deg.03)$ near the center of sgr-B2 cloud exhibits an anomalously high velocity dispersion of $\delta v \sim 40 \ekms$ around the center velocity $\vlsr\sim+72 \ekms$ (figure \ref{fig-lvd}).
Such a large velocity width cannot be explained neither by the expansion of an HII region, which is on the order of $\sim 10 \ekms$, nor by the cloud collision at colliding speed $\sim 20$--30 \kms.
This compact cloud is near to the high-velocity clump No. 138, $(l,b,\vlsr)=0\deg671, -0\deg.032,+94.0 \ekms)$ listed as a "wing" in the catalogue by \citet{oka22hvc}.
However, the center velocity is significantly different, so that their physical relation is not clear at the moment.

\subsection{Dual-side compression (DSC) of molecular cloud and high efficiency of star formation}
\label{ss-dsc}

MBS is composed of low-temperature and high-density molecular gas.
On the other hand, GCH is high-temperature and low-density HII gas expanded around OB clusters. 
Both the MBS and GCH show a similar morphology as the GCH is sheathed by MBS after their interaction and mutual deformation. 
The expansion of GCH is suppressed by MBS on the up-stream side, while it becomes more free in the tail. 
The side wall of MBS also suppresses the HII expansion, and guides the HII gas into a sharper cone. 

During their interaction, the GCH and MBS act to trigger star formation in such a way that GCH compresses the stacked molecular gas at the bow head from inside, and MBS compresses the molecular clouds from outside (figure \ref{fig-illust}). 
The supersonic flow by the spiral arm's gravitational potential convey and supply molecular clouds to the bow head. 
Thus, the bow head becomes an efficient SF site by the "dual-side dynamical compression" (DSC) by the MBS and GCH. 

\red{
The frequency for one cloud to experience the DSC is equal to the frequency for a cloud to encounter the spiral arms in the CMZ, which is given by
$f\sim (\Vrot-\Vpat)/(\pi R)$,
where $\Vrot$ is the rotation velocity, $\Vpat=\Omega_{\rm p} R$ the pattern velocity, and $\Omega_{\rm p}$ is the angular speed of the pattern.
If we assume that 
$R\sim 100$ pc,
$V_{\rm rot} \sim 150$ \kms
and the pattern's angular velocity is equal to that in the Galactic disc, 
$\Omega_{\rm p}\sim 20$ \kms kpc$^{-1}$, yielding 
$\Vpat\sim 40$ \kms, we obtain 
$f\sim 4\times 10^{-7}$ yr$^{-1}$.
}

\subsection{MBS in extra-galactic CMZ}

The bow-shock feature is ubiquitous not only in the Galaxy, but also in spiral galaxies.
In figure \ref{fig-msb-m74m83} we show a bow shock calculated in order to simulate a giant MSB and GCH observed in M83, which are well fitted by fronts with bow radii $\Rbow \sim$25--60 pc \citep{sof18}.
Similar bow shocks are observed  also in the recent JWST images of nearby spiral galaxies such as M74 \citep{2023ApJ...944L..17L,2023ApJ...944L..22B} as shown in the right panel of figure \ref{fig-msb-m74m83}.
So, we may conclude that Sgr B molecular complex is one of the typical bow-shock phenomena in disc galaxies.

\subsection{Evolution of Sgr-B molecular complex}

Finally we propose a possible scenario of evolution of the Sgr B1 and B2 clouds based on the observational facts presented in this paper.
We consider that the CMZ is an ordinary gaseous disc rotating in the Galactic potential, which is superposed by a spiral density wave or a bar potential.
The orbits of the ISM and the molecular clouds are disturbed to cause the galactic shock wave in the oval potential well.
Molecular gas and clouds entering the shock wave are strongly compressed to form stars.
Each molecular cloud is compressed not only by the increase of the pressure of the surrounding gas, but also by the stagnation at the bow-shock head, as well as by the counter-compression by the thermal pressure of the HII region expanding from the newly born OB stars.
This view is ubiquitous in the discs of the Milky Way and in spiral galaxies, and therefore, the Sgr B region can be regarded to be a normal star forming cloud.

\begin{figure}
\begin{center}     
\includegraphics[width=8.8cm]{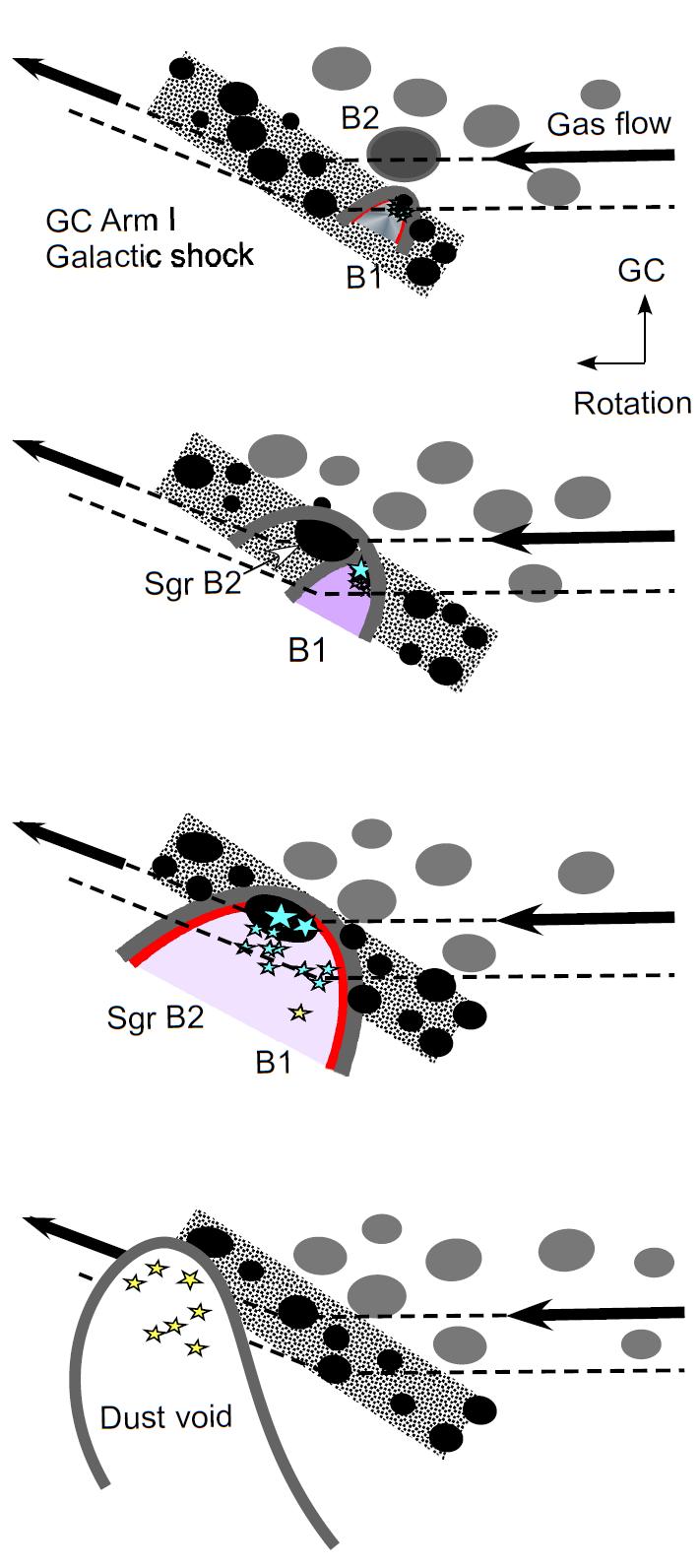}  
\end{center}
\caption{Evolution of the Sgr-B1 and B2 clouds in the GC Arm I, which is a galactic shock wave in the central oval potential.
This view is ubiquitous in the discs of the Milky Way and spiral galaxies.
This scenario will apply to the network of deformed dust voids containing SF regions as observed generally in spiral galaxies such as M74 \citep{2023ApJ...944L..17L,2023ApJ...944L..22B} (figure \ref{fig-msb-m74m83}).
}
\label{fig-evo}	
\end{figure}

Figure \ref{fig-evo} illustrates the evolution of the Sgr B1/B2 complex.
A molecular cloud orbiting around the GC enters the GC Arm I, where star formation is triggered by the galactic shock compression, and produces Sgr B1 SF region.
Then, a neighboring bigger cloud enters the shock and is compressed to form denser cloud followed by star, which creates Sgr B2, where the feedback of the prior SF in Sgr B1 enhanced the SF by the pressured of the expanding HII region.
The subsequent gas flow from behind causes a bow shock headed by the Sgr B2 cloud, whose outskirts in the down-stream side  composes the extended molecular bubble.
As the SF region and molecular bubble around Sgr B2 evolve, Sgr B1 is merged by the molecular bow of Sgr B2, which is observed at the present time.
This sequential process of the cloud compression, bow shock, star formation, SF feedback, and dual compression of the bow head is a general gas dynamical mechanism in a spiral arm of a galactic disc.
This mechanism will, therefore, also explain the origin of the star-formation belt in the entire CMZ disc \citep{sof22-3d}.
As the molecular complex expands, it is trailed toward the downstream by the Galactic rotation, attaining an elongated structure toward the Sun as well as to the east on the sky, which is indeed observed as the lopsidedness of the complex as shown in figure \ref{fig-rgb}.
The scenario would be more general and may be applied to the network of deformed dust voids containing SF regions as widely observed in spiral galaxies \citep{2023ApJ...944L..17L,2023ApJ...944L..22B} (figure \ref{fig-msb-m74m83}).
However, the vertical displacement of the whole complex by $\sim 30$ pc toward south from the Galactic plane remains an unanswered question.

\subsection{Comparison with other models}

We comment on the other two models so far proposed for the Sgr-B molecular complex.  

\citet{bal88} interpreted the "hole and ring" at $\vlsr \sim 30$ \kms as due to gravitational perturbation by the massive Sgr-B2 cloud at $\sim 60$ \kms. This idea is interesting in the sense that we may be watching a contraction phase of a giant cloud by the self gravity, because the central region contracts faster than outskirts due to shorter Jeans time. Difficulty in this model would be the difference in the systemic-velocity of $\delta \vlsr \sim 30$--40 \kms. If we consider the Galactic rotation, the mutual distance would be too far for gravitational attraction between the core and ring. In the present model, we interpret the hole as a feature due to velocity slicing at $\sim 30$--40 \kms of a conical cylinder continuously distributed in the $(l,v,\vlsr)$ space.
     
In the cloud-collision model \citep{eno22}, the 70-\kms cloud associated with Sgr B2 and the 40-\kms cloud are assumed to be two distinct unbound bodies orbiting on a straight line (extremely hyperbolic orbits) with zero angular momentum. 
In the circumstance of CMZ, the mean-free path and collision frequency of clouds of radius $\sim 10 \epc$ randomly moving at $v_{\rm c}\sim 30$ \kms are estimated to be $L_{\rm mfp}\sim 400 \epc$ and $t_{\rm c}\sim 10$ Myr, respectively.  
Since the 70-\kms cloud (Sgr B2) composes the main LVD ridge of GC Arm I rotating with the disc of CMZ, the 40-\kms cloud would be the passerby cloud.
If so, the 40-\kms cloud had to come from the far side of Sgr B2, flying in the counter sense to the Galactic rotation.       
 
A concern about this model is the assumed zero angular momentum.  
In order to prove such a special (straight) orbit in order for a head-on collision of two unbound objects, determination of the six orbital elements is the inevitably necessary condition, which, however, seems to have been not obtained.  

Another concern is that the collision model ignores the rapid Galactic rotation in the deep gravitational potential of the Galactic Centre, which prohibits such a straight or hyperbolic orbit.
In the rotating system, the angular velocity 
$ \Omega=\vrot/R$
yields the Coriolis force (acceleration)  
$ a_{\rm cor}= -2 \Omega \times v_{\rm c} $ 
on the object moving at velocity $v_{\rm c}$ with respect to the rotating coordinates.
The object therefore attains the epicyclic motion, drawing a local circle of radius
$ R_{\rm epi}=v_{\rm c}/(2\Omega_{\rm})$
with a period of
$ P_{\rm epi}= \pi/\Omega.$
For the 40-\kms cloud having $\vrot\sim 150$ \kms, $R\sim 100$ pc, and $v_{\rm c}\sim 30$ \kms, we obtain 
$R_{\rm epi} \sim  10$ pc and
$P_{\rm epi}\sim 2$ Myr.
Namely, the 40-\kms cloud is orbiting around the epicyclic center 10 pc displaced from its present position.
Since Sgr B2 is rotating at the same angular speed ($\Omega$) around the GC, the 40-\kms cloud encounter/passes Sgr B2 every $P_{\rm epi}\sim 2$ My during its epicyclic oscillation.  

Such difficulty does not exist in our bubble model, which attributes the whole Sgr-B complex to a large single system rotating with the disc of CMZ.  
The continuous distribution of gas is indeed confirmed in the channel maps (figure \ref{fig-chan+mkat}), position-velocity diagrams (figure \ref{fig-lvd}), and in the FOT maps (figure \ref{fig-3d-xyz})


\section{Summary}
\label{sec6}

We revisited the expanding-bubble model for the Sgr B molecular complex \citep{sof90} based on the analysis of high-resolution CO-line cube data of the Galactic Centre observed with the Nobeyama 45-m radio telescope by \citet{tok19}.
Considering the rotation of the CMZ at $\vrot\sim 150 \ekms$ and applying the FOT method to the CO-line cube, we derived the 3D structure of Sgr B molecular complex.
The 3D molecular-gas distribution exhibited a conical-horn structure with the densest Sgr-B2 cloud at $\vlsr\sim 70$ \kms in the farthest end followed by the conical bubble at $\vlsr \sim 60$--20 \kms open toward the Sun.
The entire Sgr B complex is elongated in the direction of the Sun with full 3D sizes of 
$\Delta X \times \Delta Y \times \Delta Z \sim 40 \epc \times 100 \epc \times 20 \epc$, where $Z$ is the axis toward increasing longitude, $Y$ is the axis toward the Sun, and $Z$ toward the north. 

We proposed a formation mechanism of the derived 3D structure of Sgr B complex as due to a molecular  bow shock associated with a giant cometary HII region in the galactic shock wave in the GC Arm I of the CMZ.
We further argued that dual-side compression (DSC) of the molecular cloud by the Galactic shock wave and expanding HII region acts to enhance the star formation in the GC Arm I, resulting in the active SF regions Sgr B1 and B2, whose feedback produced the conical-horn shape of the molecular complex.
We conclude that the star formation in the GC is triggered by the Galactic-scale shock waves propagating in the dense CMZ as well as by the feedback through expansion of HII regions. 

\section*{Acknowledgements} 
The data analysis was performed at the Astronomical Data Analysis Center of the National Astr. Obs. of Japan.
The author thanks Prof. Tomoharu Oka for the Nobeyama CO-line survey of GC and Dr, I. Heywood for the MeerKAT archival image.
 
\section*{Data availability} 
The CO data were taken from
https:// www.nro.nao.ac.jp/ $\sim$nro45mrt/html/ results/data.html. 
The MeerKAT radio data were downloaded from https:// archive-gw-1.kat.ac.za/ public/ repository/ 10.48479/fyst-hj47/index.html. 
The optical image of M83 was reproduced from the web sites of STSci at 
http://www.stsci.edu/hst/wfc3/ and NASA at https://apod.nasa.gov/apod/.
The infrared image of M74 was taken from url:
https://www.esa.int/
Science\_Exploration/
Space\_Science/Webb/
Webb\_inspects \_the\_heart
\_of\_the \_Phantom\_Galaxy  

\section*{Conflict of interest}
The author declares that there is no conflict of interest.


    
\end{document}

%% file: define.tex
\def\be{\begin{equation}}\def\ee{\end{equation}}\def\vlsr{v_{\rm LSR}} \def\Msun{M_\odot} \def\deg{^\circ}  
\def\vrot{V_{\rm rot}} \def\Vrot{\vrot}\def\co{$^{12}$CO } \def\coth{$^{13}$CO } \def\Xco{X_{\rm {^{12}}CO}}   \def\Tb{T_{\rm B}} \def\Htwo{H$_2$ }      \def\kms{km s$^{-1}$} \def\Ico{I_{\rm CO}}  \def\mH{m_{\rm H}}  \def\Ico{I_{\rm ^{12}CO}}  
 \def\htwo{H$_2$} \def\Tb{T_{\rm B}}   \def\mH{m_{\rm H}} \def\ekms{{\rm \ km \ s^{-1}}}  \def\epc{{\rm \ pc} }   \def\apj{ApJ} \def\aap{AA} \def\mnras{MNRAS} \def\pasj{PASJ} \def\aj{AJ} \def\xcounit{H$_2$ cm $^{-2}$ [K km s$^{-1}]^{-1}$}         
\def\Msun{M_{\odot \hskip-5.2pt \bullet}}    \def\kms{km s$^{-1}$}  \def\deg{^\circ}   \def\Htwo{H$_2$\ }      
     \def\bc{\begin{center}}\def\ec{\end{center}} 
\def\xcounit{\Htwo cm$^{-2}$ [K \kms]}   
  
   \def\Rbow{R_{\rm bow}}   \def\Lsun{L_\odot}\def\Rhii{R_{\rm HII}} \def\nuv{n_{\rm UV}} \def\ni{n_{\rm i}} \def\ne{n_{\rm e}}\def\ar{a_{\rm r}}
\def\Te{T_{\rm e}}\def\Tn{T_{\rm n}}\def\({\left(} \def\){\right)}\def\[{\left[} \def\]{\right]}  \def\L{\mathcal{L}}     \def\vpat{V_{\rm pattern}}\def\red{\textcolor{red}} 
\def\Vrot{V_{\rm rot}}\def\Vpat{V_{\rm pat}}

%% file: MS.bbl
\begin{thebibliography}{}
 

\bibitem[Arce \& Goodman(2002)]{2002ApJ...575..928A} Arce, H.~G., \& Goodman, A.~A.\ 2002, \apj, 575, 928 



\bibitem[Arimoto et al.(1996)]{ari96}
Arimoto, N., Sofue, Y., \& Tsujimoto, T.\ 1996, \pasj, 48, 275. 

\bibitem[Arthur \& Hoare(2006)]{2006ApJS..165..283A} Arthur, S.~J., \& Hoare, M.~G.\ 2006, \apjs, 165, 283 

    \bibitem[Baba et al.(2017)]{baba17}
    Baba, J., Morokuma-Matsui, K., \& Saitoh, T.~R.\ 2017, \mnras, 464, 246.
    
\bibitem[Bally et al.(1987)]{1987ApJS...65...13B} Bally, J., Stark, A.~A., Wilson, R.~W., et al.\ 1987, \apjs, 65, 13. 

\bibitem[Bally et al.(1988)]{bal88}
Bally, J., Stark, A.~A., Wilson, R.~W., et al.\ 1988, \apj, 324, 223. 

\bibitem[Barnes et al.(2023)]{2023ApJ...944L..22B}
Barnes, A.~T., Watkins, E.~J., Meidt, S.~E., et al.\ 2023, \apjl, 944, L22. 

\bibitem[Bolatto et al.(2013)]{2013ARA&A..51..207B} Bolatto, A.~D., Wolfire, M., \& Leroy, A.~K.\ 2013, \araa, 51, 207   
  
 
 
\bibitem[Deharveng et al.(2015)]{2015A&A...582A...1D} Deharveng, L., Zavagno, A., Samal, M.~R., et al.\ 2015, \aap, 582, A1 

 \bibitem[Downes et al.(1980)]{dow80}
 Downes, D., Wilson, T.~L., Bieging, J., et al.\ 1980, \aaps, 40, 379
 
\bibitem[Dyson(1975)]{1975Ap&SS..35..299D} Dyson, J.~E.\ 1975, \apss, 35, 299 
 
\bibitem[Enokiya \& Fukui(2022)]{eno22}
Enokiya, R. \& Fukui, Y.\ 2022, \apj, 931, 155. 

\bibitem[Fujimoto(1968)]{fuj68} Fujimoto, M. 1968, in IAU Symp. 29, Nonstable Phenomena in Galaxies, ed. M. Arakeljan (Yerevan: Publishing House of the Academy of Sciences of Armenian SSR), 453

\bibitem[Fukui et al.(2021)]{fuk21}
Fukui, Y., Habe, A., Inoue, T., et al.\ 2021, \pasj, 73, S1. 

\bibitem[Fukuda \& Hanawa(2000)]{2000ApJ...533..911F} Fukuda, N., \& Hanawa, T.\ 2000, \apj, 533, 911  

\bibitem[GRAVITY Collaboration et al.(2019)]{2019A&A...625L..10G} GRAVITY Collaboration, Abuter, R., Amorim, A., et al.\ 2019, \aap, 625, L10. 


\bibitem[Hasegawa et al.(1994)]{has94}
Hasegawa, T., Sato, F., Whiteoak, J.~B., et al.\ 1994, \apjl, 429, L77. 

\bibitem[Henshaw et al.(2023)]{2023ASPC..534...83H} Henshaw, J.~D., Barnes, A.~T., Battersby, C., et al.\ 2023, Protostars and Planets VII, 534, 83. 

\bibitem[Heywood et al.(2022)]{hey22}
Heywood, I., Rammala, I., Camilo, F., et al.\ 2022, \apj, 925, 165. 

\bibitem[Kruijssen et al.(2019)]{kru19}
Kruijssen, J.~M.~D., Dale, J.~E., Longmore, S.~N., et al.\ 2019, \mnras, 484, 5734. 

\bibitem[Lee et al.(2023)]{2023ApJ...944L..17L} Lee, J.~C., Sandstrom, K.~M., Leroy, A.~K., et al.\ 2023, \apjl, 944, L17. 
 
\bibitem[Liu et al.(2013)]{2013ApJ...778L..41L} Liu, G., Calzetti, D., Hong, S., et al.\ 2013, \apjl, 778, L41 
  

\bibitem[McQuinn et al.(2017)]{2017AJ....154...51M} McQuinn, K.~B.~W., Skillman, E.~D., Dolphin, A.~E., et al.\ 2017, \aj, 154, 51. 

\bibitem[Mehringer et al.(1993)]{meh93}
Mehringer, D.~M., Palmer, P., Goss, W.~M., et al.\ 1993, \apj, 412, 684. 
 
\bibitem[Mehringer et al.(1992)]{meh92}
Mehringer, D.~M., Yusef-Zadeh, F., Palmer, P., et al.\ 1992, \apj, 401, 168. 


\bibitem[Ogura(1995)]{1995Ap&SS.224..151O} Ogura, K.\ 1995, \apss, 224, 151 
  
\bibitem[Oka et al.(1998)]{oka98}
Oka, T., Hasegawa, T., Sato, F., et al.\ 1998, \apjs, 118, 455. 

\bibitem[Oka et al.(2001)]{oka01}
Oka, T., Hasegawa, T., Sato, F., et al.\ 2001, \pasj, 53, 787. 

\bibitem[Oka et al.(2022)]{oka22hvc}
Oka, T., Uruno, A., Enokiya, R., et al.\ 2022, \apjs, 261, 13. 

\bibitem[Reipurth et al.(2002)]{2002AJ....123..362R} Reipurth, B., Heathcote, S., Morse, J., Hartigan, P., \& Bally, J.\ 2002, \aj, 123, 362  

\bibitem[Roberts(1969)]{rob69}
Roberts, W.~W.\ 1969, \apj, 158, 123. 

\bibitem[Roberts(1972)]{rob72}
Roberts, W.~W.\ 1972, \apj, 173, 259. 

\bibitem[Roberts et al.(1979)]{rob79}
Roberts, W.~W., Huntley, J.~M., \& van Albada, G.~D.\ 1979, \apj, 233, 67. 


\bibitem[Sakashita(1971)]{sak71} Sakashita, S.\ 1971, ApSS, 14, 431

\bibitem[Sato et al.(2000)]{sat00}
Sato, F., Hasegawa, T., Whiteoak, J.~B., et al.\ 2000, \apj, 535, 857. 
 
\bibitem[Sofue(1985)]{1985PASJ...37..507S} Sofue, Y.\ 1985, \pasj, 37, 507
 
\bibitem[Sofue(1990)]{sof90}
Sofue, Y.\ 1990, \pasj, 42, 827

\bibitem[Sofue(1995)]{sof95}
Sofue, Y.\ 1995, \pasj, 47, 551. 

\bibitem[Sofue(2013)]{sof13}Sofue, Y.\ 2013, \pasj, 65, 118. 
 
\bibitem[Sofue(2018)]{sof18}
Sofue, Y.\ 2018, \pasj, 70, 106. 
 
\bibitem[Sofue(2019)]{2019PASJ...71..104S} Sofue, Y.\ 2019, \pasj, 71, 104. 

\bibitem[Sofue(2022)]{sof22-3d}
Sofue, Y.\ 2022, \mnras, 516, 907. 

\bibitem[Sofue(2023)]{sof23g18}
Sofue, Y.\ 2023, \mnras, 525, 4540. 


\bibitem[Sofue et al.(2019)]{2019PASJ...71S...1S} Sofue, Y., Kohno, M., Torii, K., et al.\ 2019, \pasj, 71, S1. 
 
\bibitem[Sorensen et al.(1976)]{soren76}
Sorensen, S.-A., Matsuda, T., \& Fujimoto, M.\ 1976, \apss, 43, 491. 

 \bibitem[Sormani et al.(2015)]{sorma15}
 Sormani, M.~C., Binney, J., \& Magorrian, J.\ 2015, \mnras, 451, 3437. 
\bibitem[Sormani et al.(2020)]{sorma20}
Sormani, M.~C., Tress, R.~G., Glover, S.~C.~O., et al.\ 2020, \mnras, 497, 5024.
 
\bibitem[Steggles et al.(2017)]{2017MNRAS.466.4573S} Steggles, H.~G., Hoare, M.~G., \& Pittard, J.~M.\ 2017, \mnras, 466, 4573 

\bibitem[Tanaka et al.(2018)]{2018ApJS..236...40T} Tanaka, K., Nagai, M., Kamegai, K., et al.\ 2018, \apjs, 236, 40. 

\bibitem[Tokuyama et al.(2019)]{tok19}
Tokuyama, S., Oka, T., Takekawa, S., et al.\ 2019, \pasj, 71, S19. 

\bibitem[Tsuboi et al.(1997)]{tsu97}
Tsuboi, M., Ukita, N., \& Handa, T.\ 1997, \apj, 481, 263.

\bibitem[Tsuboi et al.(2009)]{tsu09}
Tsuboi, M., Miyazaki, A., \& Okumura, S.~K.\ 2009, \pasj, 61, 29. 

\bibitem[Tsuboi et al.(1999)]{1999ApJS..120....1T} Tsuboi, M., Handa, T., \& Ukita, N.\ 1999, \apjs, 120, 1. 

\bibitem[Tsuboi et al.(2015)]{tsu15}
Tsuboi, M., Miyazaki, A., \& Uehara, K.\ 2015, \pasj, 67, 90. 
   
\bibitem[van Buren et al.(1990)]{1990ApJ...353..570V} van Buren, D., Mac Low, M.-M., Wood, D.~O.~S., \& Churchwell, E.\ 1990, \apj, 353, 570  

\bibitem[Vink(2022)]{vin22}
Vink, J.~S.\ 2022, \araa, 60, 203. 

\bibitem[Wada et al.(2011)]{wad11}
Wada, K., Baba, J., \& Saitoh, T.~R.\ 2011, \apj, 735, 1. 

\bibitem[Whitmore et al.(2011)]{2011ApJ...729...78W} Whitmore, B.~C., Chandar, R., Kim, H., et al.\ 2011, \apj, 729, 78. 

\bibitem[Wilkin(1996)]{wil96}
Wilkin, F.~P.\ 1996, \apjl, 459, L31. 

\bibitem[Yusef-Zadeh et al.(2022)]{yus22a}
Yusef-Zadeh, F., Arendt, R.~G., Wardle, M., et al.\ 2022, \apjl, 925, L18. 

\bibitem[Zhu et al.(2015)]{zhu15}
Zhu, F.-Y., Zhu, Q.-F., Li, J., et al.\ 2015, \apj, 812, 87. 
 

\end{thebibliography}
